\documentclass[apj]{emulateapj}
\usepackage{graphicx}
\usepackage{psfig}

\shorttitle{Testing Atomic Data for SDO}
\shortauthors{Testa et al.}
\usepackage{color} 
\usepackage{amsmath}           
\usepackage{amsfonts}

\def \arcsec {\hbox{$^{\prime\prime}$}}

\def \lx   {$L_{\rm X}$}

\def \ll   {$\lambda$}
\def \ergs {erg~s$^{-1}$}
\def \flxu {ph~cm$^{-2}$~s$^{-1}$}

\def \sdo   {{\em SDO}}
\def \cha   {{\em Chandra}}
\def \xmm   {{\em XMM-Newton}}

\def \cv   {C\,{\sc v}}
\def \cvi   {C\,{\sc vi}}
\def \nvi   {N\,{\sc vi}}
\def \nvii   {N\,{\sc vii}}
\def \nevii   {Ne\,{\sc vii}}
\def \neviii   {Ne\,{\sc viii}}
\def \neix   {Ne\,{\sc ix}}
\def \nex   {Ne\,{\sc x}}

\def \ovii   {O\,{\sc vii}}
\def \oviii   {O\,{\sc viii}}
\def \fevii   {Fe\,{\sc vii}}
\def \feviii  {Fe\,{\sc viii}}
\def \feix    {Fe\,{\sc ix}}
\def \fex     {Fe\,{\sc x}}

\def \fexii   {Fe\,{\sc xii}}
\def \fexiii  {Fe\,{\sc xiii}}
\def \fexiv   {Fe\,{\sc xiv}}

\def \fexvi   {Fe\,{\sc xvi}}
\def \fexvii  {Fe\,{\sc xvii}}
\def \fexviii {Fe\,{\sc xviii}}

\def \nixi  {Ni\,{\sc xi}}
\def \nixii  {Ni\,{\sc xii}}
\def \nixiii  {Ni\,{\sc xiii}}

\def \svii   {S\,{\sc vii}}
\def \s9     {S\,{\sc ix}}
\def \sx     {S\,{\sc x}}
\def \sxi     {S\,{\sc xi}}

\def \siviii   {Si\,{\sc viii}}
\def \siix     {Si\,{\sc ix}}
\def \six     {Si\,{\sc x}}
\def \sixi     {Si\,{\sc xi}}
\def \sixii     {Si\,{\sc xii}}

\def \caxi   {Ca\,{\sc xi}}
\def \heii   {He\,{\sc ii}}
\def \mgix     {Mg\,{\sc ix}}

\def\ion[#1 #2]{#1\,{\sc #2}}
\def\dens[#1]{10$^{#1}$\hskip 1.5pt{cm$^{-3}$}}
\def\densr[#1 #2]{10$^{#1}$\hskip 1pt{--}\hskip .5pt{10$^{#2}$}\hskip 1.5pt{cm$^{-3}$}}
\def\fl[#1 #2]{{#1}$\pm${#2}}
\def\orb[#1 #2]{{$#1^{#2}$}}
\def\ls[#1 #2]{{$^{#1}${#2}}}
\def\tm[#1 #2 #3]{{$^{#1}${#2}$_{#3}$}}

\newcounter{ion}

\begin{document}
\title{Testing EUV/X-ray Atomic Data for the {\em Solar Dynamics Observatory}}
\author{Paola Testa\altaffilmark{1}, Jeremy J.\ Drake \altaffilmark{1}, 
Enrico Landi\altaffilmark{2}}
\altaffiltext{1}{Smithsonian Astrophysical Observatory,
	60 Garden street, MS 58, Cambridge, MA 02138, USA; 
	ptesta@cfa.harvard.edu}
\altaffiltext{2}{Department of Atmospheric, Oceanic and Space
  Sciences, University of Michigan 2455 Hayward St., Ann Arbor MI
  48109 USA}

\begin{abstract}

The Atmospheric Imaging Assembly (AIA) and the Exteme-ultraviolet
Variability Experiment (EVE) onboard the Solar Dynamics Observatory include
spectral windows in the X-ray/EUV band. 
Accuracy and completeness of the atomic data in this wavelength range 
is essential for interpretation of the spectrum and irradiance of the solar 
corona, and of SDO observations made with the AIA and EVE instruments.
Here we test the X-ray/EUV data in the CHIANTI database to assess their 
completeness and accuracy in the \sdo\ bands, with particular focus on 
the 94\AA\ and 131\AA\ AIA passbands.  
Given the paucity of solar observations adequate for this
purpose, we use high-resolution X-ray spectra of the low-activity
solar-like corona of Procyon obtained with the \cha\ Low Energy
Transmission Grating Spectrometer (LETGS). 
We find that while spectral models overall can reproduce quite well the
observed spectra in the soft X-ray range $\lambda \lesssim 50$\AA, and 
at the EUV wavelengths $\lambda \gtrsim 130$\AA, they significantly 
underestimate the observed flux in the 50-130\AA\ wavelength range. 
The model underestimates the observed flux by a variable factor ranging 
from $\approx 1.5$, at short wavelengths below $\sim$50\AA, up to 
$\approx$5-7 in the $\sim 70-125$\AA\ range.
In the AIA bands covered by LETGS, i.e.\ 94\AA\ and 131\AA, we find that the 
observed flux can be underestimated by large factors ($\sim 3$ and $\sim 1.9$ 
respectively, for the case of Procyon presented here). We discuss the consequences
for analysis of AIA data and possible empirical corrections to the AIA responses 
to model more realistically the coronal emission in these passbands.  

\end{abstract}


\section{Introduction}
\label{s:intro}

The solar EUV and X-ray radiation plays a double role in the physics of the solar upper
atmosphere: by representing an important term in the energy equation, and by bearing the
signatures of  the most important physical phenomena that occur during solar activity, namely,  
flares and coronal mass ejections (CMEs). It is also one of the primary energy inputs to the Earth's upper 
atmosphere: it heats the thermosphere, creates the ionosphere, and drives a number of 
dynamical motions and photochemical reactions at different heights.  By its own nature, the 
EUV and X-ray emission of the Sun is highly variable on all time scales, from a factor 
two to several orders of magnitude \citep{Woods04}.  The largest variations 
occur at the shortest wavelengths.

The 10-170~\AA\ region includes a large number of spectral lines emitted by highly ionized
species formed at temperatures at, or above, 1~MK. These lines dominate the EUV solar
irradiance, and are of crucial importance for investigating the interaction between the 
solar radiative output and the Earth's upper atmosphere.  For example, \cite{Pawlowski08} 
showed that flare radiation, dominated by highly ionized Fe transitions 
in the 90-140~\AA\ range, can increase the density of the Earth's thermosphere by as 
much as $\simeq$15\% in less than two hours. These lines also provide excellent 
diagnostic tools for measuring the physical properties of the emitting 
plasmas under quiescent, active and flaring conditions. Furthermore, key ions 
routinely detected by {\em in-situ} measurements of the solar wind (\ion[O vii, viii], 
\ion[C v, vi]) emit very strong lines between 18~\AA\ and 40~\AA.

In recent years, the 10-170~\AA\ range has attracted considerable attention because 
of several instruments, both already launched and being developed, that can observe 
astrophysical objects in this spectral interval.  
{\em EUVE}, \cha\ and \xmm\ obtained high-resolution spectra from all kinds of 
astrophysical objects, while the Thermosphere Ionosphere Mesosphere Energetics and 
Dynamics and Solar EUV Experiment (TIMED/SEE) and the recently launched
Solar Dynamic Observatory (\sdo) include instrumentation aimed at studying the solar
corona, the solar irradiance and their variability using this range.  In particular, the 
Extreme Ultraviolet Variability Experiment (EVE,
\citealt{Woods10}) and the Atmospheric Imaging Assembly (AIA, \citealt{Lemen11,Boerner11})
on board \sdo\ include spectral windows in the 10-170~\AA\ range aimed at studying the energy 
input, storage and release mechanisms that rule coronal heating and the variability 
of the solar spectrum. AIA is a suite of 4 telescopes providing high cadence 
($\sim 12$~s, for the standard observing mode, though higher cadence is possible)
full Sun observations in 7 EUV narrow passbands, at high spatial resolution 
($\sim 0.6$\arcsec/pixel): six EUV narrowbands are centered around 94\AA, 131\AA, 
171\AA, 195\AA, 211\AA, 335\AA, respectively, which are generally dominated by 
emission of plasma at temperatures $\log (T[K]) \gtrsim 5.7$. AIA also observes 
cooler plasma in the 304\AA\ channel (dominated by \heii\ emission) and in a 
UV channel (where three passbands can be selected at $\lambda$ 1600\AA, 1700\AA,
4500\AA).  EVE measures the full disk solar irradiance in the EUV and soft X-ray 
energy range (from 1 to 1050\AA), with a cadence of 10s. EVE spectral resolution 
is $\sim 1$\AA\ between 50 and 1050\AA\ (with the {\em Multiple EUV Grating 
Spectrographs} MEGS-A and MEGS-B), and 10\AA\ in the 1-50\AA\ range (with MEGS-SAM). 
EVE also includes an EUV Spectrophotometer (ESP) measuring the irradiance in 
broad bands in the 1-390\AA\ range, and a MEGS-Photometer (MEGS-P) measuring the 
hydrogen Lyman-$\alpha$ line at 1216\AA.

In order to understand the solar EUV and X-ray spectrum, as well as using it for diagnostic
purposes, theoretical models of the sources in the solar atmosphere  need to be combined 
with spectral models (such as CHIANTI - \citealt{chianti,chianti6} - and APEC - 
\citealt{Smith01,atomdb2}) that can compute it.  A key issue in such 
modeling is the accuracy and completeness of the spectral models: inaccurate or incomplete 
sets of atomic data and transition rates can jeopardize the success of any modeling 
effort. Also, the narrow-band filters in the AIA instrument can be used for quantitative
scientific purposes only if the spectrum in the wavelength range they cover is known
with accuracy.  For this reason, available spectral models need to be benchmarked with 
observations.  Benchmarking spectral codes by comparison with X-ray (1-20~\AA, 
\citealt{Phillips99,LandiPhillips06}), EUV (170-630~\AA, \citealt{Young98, Landi02a}) 
and UV (500-1600~\AA, \citealt{Landi02}) high-resolution spectra revealed some 
discrepancies and led to substantial improvements in the available data.  
Benchmark studies focusing on the 20-170~\AA\ range are limited, such as,
for instance, the work by \cite{DelZanna09} who carried out a detailed benchmark 
of X-ray and EUV \fexvii\ lines only, or \cite{Liang10} who used \cha/LETGS 
Procyon spectra for a comparison of their \fevii-\fexvi\ atomic data with 
observed spectral lines in the 49-106\AA\ range.

The aim of the present series of papers is to test the CHIANTI atomic 
data in the 10-170\AA\ wavelength range, which is relevant to \sdo\ (AIA 
and EVE) observations.  In this paper, the first of the series, we focus on 
issues in the wavelength ranges of the two shortest-wavelength 
AIA channels, centered at 94\AA\ and 131\AA.
We also provide a broad overview of the shortcomings of CHIANTI in reproducing
observed spectra in the 10-170\AA, consequential for EVE observations. 
In the next two papers of the series, we will present a systematic 
benchmark of the CHIANTI data with the observed lines, ion by ion (Drake 
et al.\ 2011 and Landi et al.\ 2011, in preparation). 

Despite the observational attention given to the 10-170~\AA\ wavelength range, in 
the context of atomic physics it has been somewhat neglected in the recent literature.  
For the last two decades or so, the atomic data 
effort stemming from solar physics has been directed  toward the wavelength
ranges covered by the high-resolution spectrometers on board the Solar and Heliospheric 
Observatory ({\em SoHO}) and {\em Hinode}, all exceeding 170~\AA. Also, few high resolution spectra
have ever been recorded from the Sun in this range, all of them 25 or more years ago, 
since most of the rocket- and satellite-borne instrumentation built in the past was 
optimized to work either below 20~\AA, or above 170~\AA. Ironically, the best spectra
in the 10-170~\AA\ wavelength range available today have been observed from much fainter
stellar sources with the \cha\ Low Energy Transmission Grating Spectrometer (LETGS; \citealt{letgs}) 
and, at lower resolution, with the Extreme Ultraviolet Explorer ({\em EUVE}; \citealt{EUVE}). 
In order to test CHIANTI, we chose to use X-ray/EUV spectra of the low-activity 
solar-like coronal emission of the subgiant Procyon, observed with \cha/LETGS, which is
characterized by spectral resolution ($\Delta \lambda \sim 0.05$\AA) significantly better 
than the resolution of EVE ($\Delta \lambda \sim 1$\AA). Procyon (F5~IV; 
$\alpha$~CMi, HD~61421; d=3.51~pc, \citealt{vanLeeuwen07}) is one of the brighest 
stars in the sky and also thanks to its proximity has been very well studied at 
optical to X-ray energies \citep[e.g.,][]{Steffen85,Drake95,Allende02}. The coronal 
emission of Procyon has been studied in detail in the past three decades at EUV and 
X-ray wavelength with several instruments \citep[e.g.,][]{Lemen89,Drake95,Raassen02}, 
indicating that the X-ray emission is rather constant (\lx$\sim 2 \times 10^{28}$~\ergs, 
in the energy range 0.1-2.4~keV based on the {\em ROSAT} All-Sky Survey; \citealt{Huensch99}), and 
characterized by a relatively cool plasma thermal distribution that peaks around 1-3~MK. 
These temperatures are close to values typical of non-flaring solar plasmas, therefore 
making Procyon an excellent X-ray source to benchmark the atomic data for plasma conditions 
typically observed by \sdo.

The observations are described in Section~\ref{s:obs}. The data analysis and
results of the determination of the plasma temperature distribution are
presented and discussed in Section~\ref{ss:results}. We summarize our
findings and draw our conclusions in Section~\ref{s:conclusions}.

\begin{figure*}[!ht]\vspace{-0.1cm}
\centerline{\includegraphics[scale=0.7,angle=90]{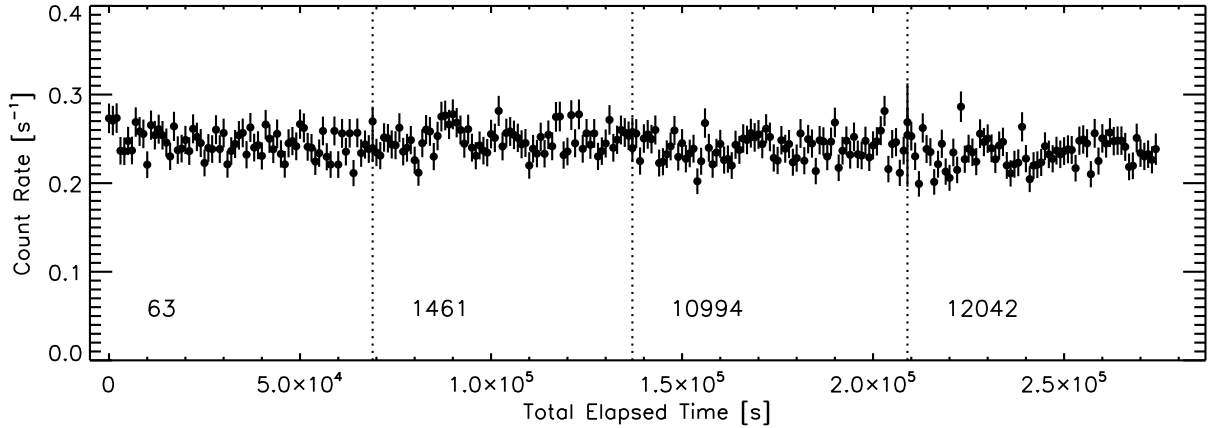}\vspace{-6cm}}
\caption{Light curves derived for each pointing from the 0th order photon
 events, binned at an interval of 1ks. \label{fig:lc}} 
\end{figure*}

\section{Observations and Data Reduction}
\label{s:obs}

We analysed \cha\ spectra of Procyon
obtained in four different pointings (two in November 1999 and two in
December 2009), using the LETG
and High Resolution Camera spectroscopic detector (HRC-S) in its
standard instrument configuration. 
Observational data were obtained from the public \cha\ Data Archive,
and reduced using the CIAO software package (v4.3).
For the analysis presented in this paper, we used the Package for
Interactive Analysis of Line Emission (PINTofALE, \citealt{PoA}). 

\begin{deluxetable}{rlccc} 
\footnotesize
\tablecolumns{5} 
\tablewidth{0pc} 
\tablecaption{Summary of \cha\ LETG+HRC-S observations of Procyon used for the analysis.
	\label{tab:obs}} 
\tablehead{ 
\colhead{Obs.}  & \colhead{Start (UT)} & \colhead{exp.} & 
\colhead{F$_{\rm X}$\tablenotemark{a}}  & 
\colhead{L$_{\rm X}$\tablenotemark{b}}  \\
\colhead{ID} & \colhead{} & \colhead{[ks]} & \colhead{[erg~cm$^{2}$~s$^{-1}$]} & \colhead{[erg~s$^{-1}$]}
}
\startdata 
  
63       &  1999-11-06 21:11:32   &  69.7 & $1.95 \times 10^{-11}$ &  $3.03 \times 10^{28}$ \\
1461   &  1999-11-07 16:59:48   &  69.8 & $1.98 \times 10^{-11}$ &  $3.06 \times 10^{28}$  \\
10994 &  2009-12-15 22:05:38   &  71.3 & $2.02 \times 10^{-11}$ &  $3.15 \times 10^{28}$  \\
12042 &  2009-12-26 01:23:39   &  65.1 & $2.01 \times 10^{-11}$ &  $3.13 \times 10^{28}$  

\enddata 
\tablenotetext{a}{ X-ray flux in the range 5-165\AA\  ($\approx 0.075-2.5$~keV). }
\tablenotetext{b}{ X-ray luminosity, in the range 5-165\AA, corrected
  for absorption (assuming N$_{\rm H} = 1.15 \times 10^{18}$
  cm$^{-2}$; \citealt{Linsky95}). } 
\end{deluxetable}

\begin{figure*}[!ht]\vspace{-0.5cm}
\centerline{\includegraphics[scale=0.55,angle=90]{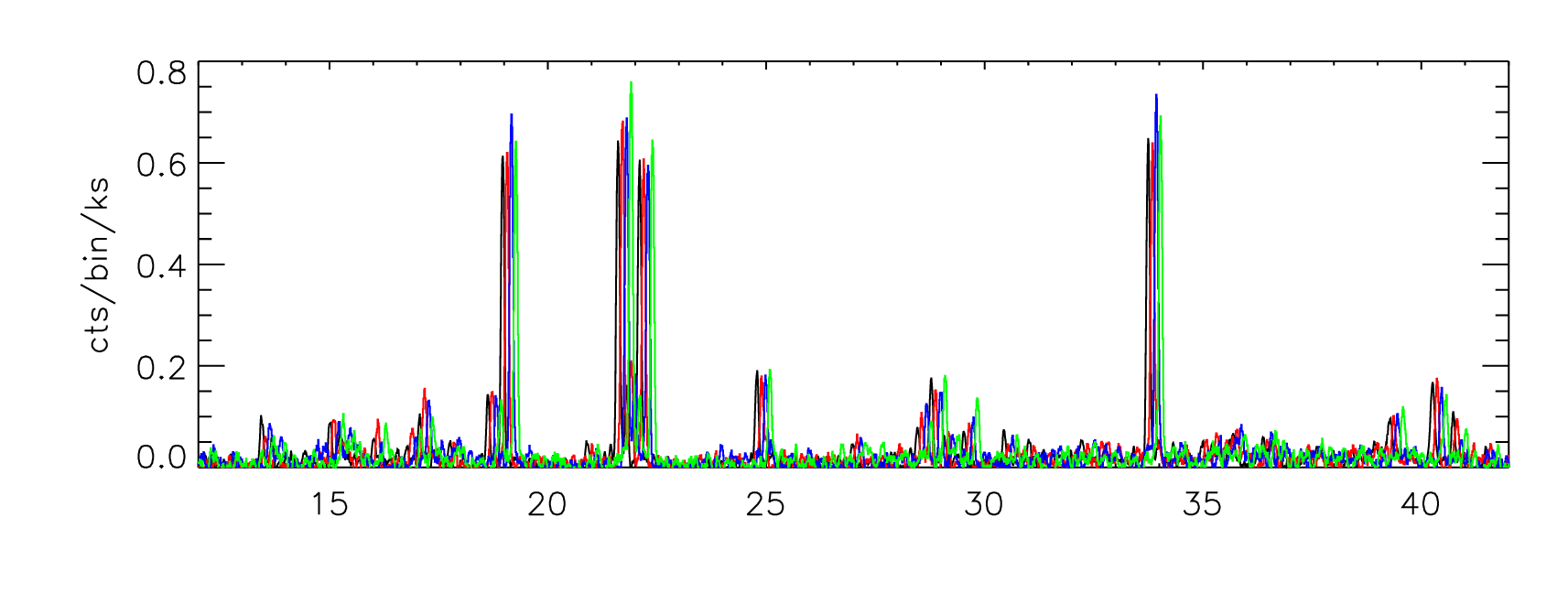}\vspace{-1.1cm}}
\centerline{\includegraphics[scale=0.55,angle=90]{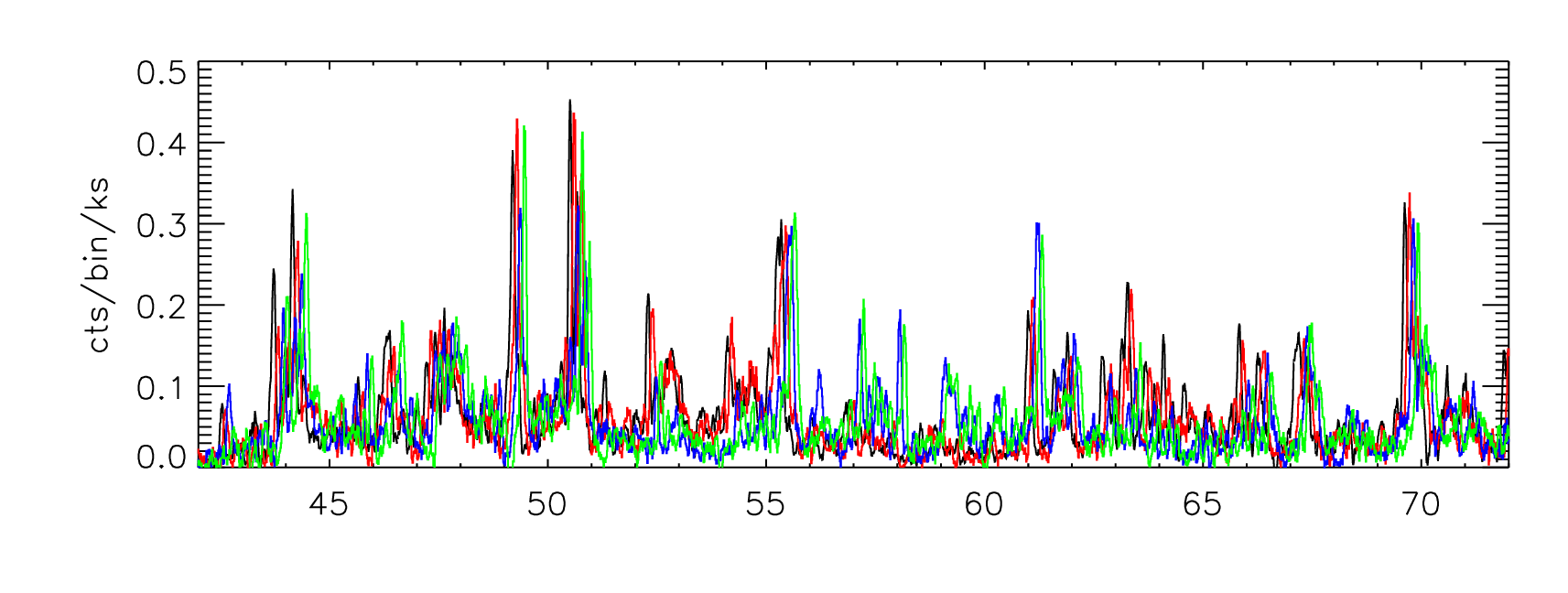}\vspace{-1.1cm}}
\centerline{\includegraphics[scale=0.55,angle=90]{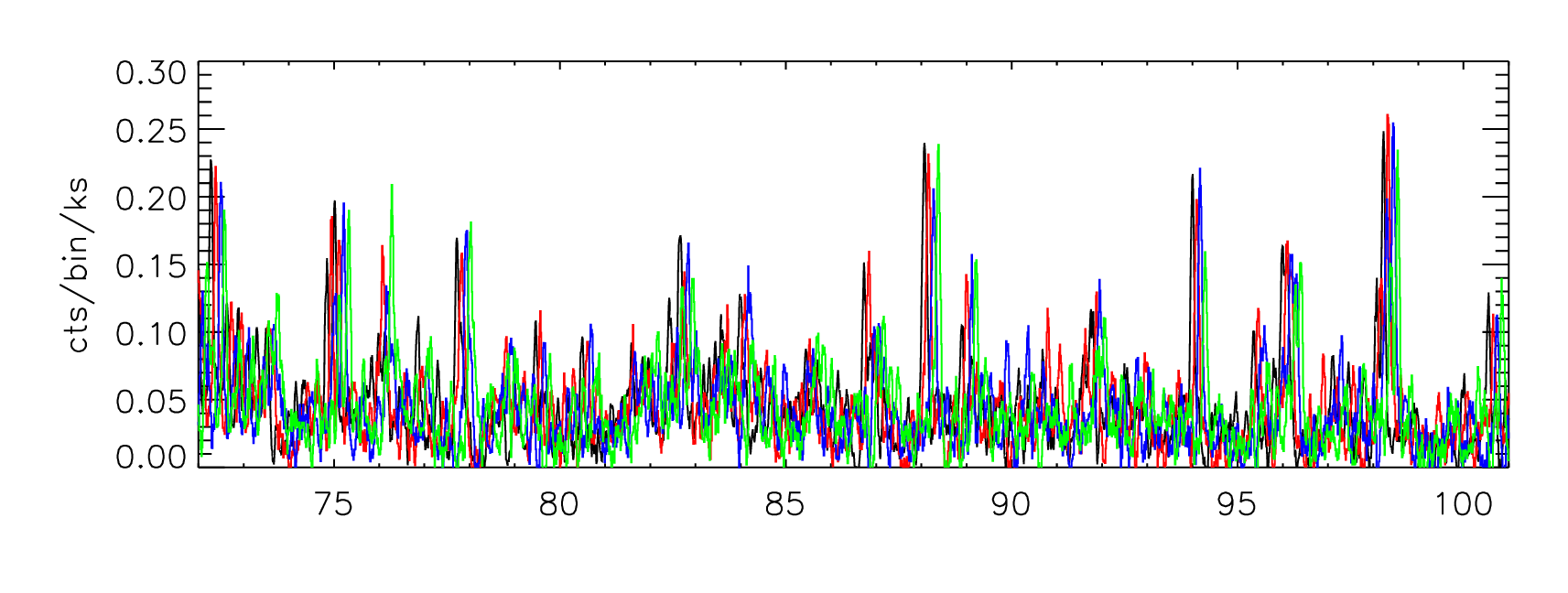}\vspace{-1.1cm}}
\centerline{\includegraphics[scale=0.55,angle=90]{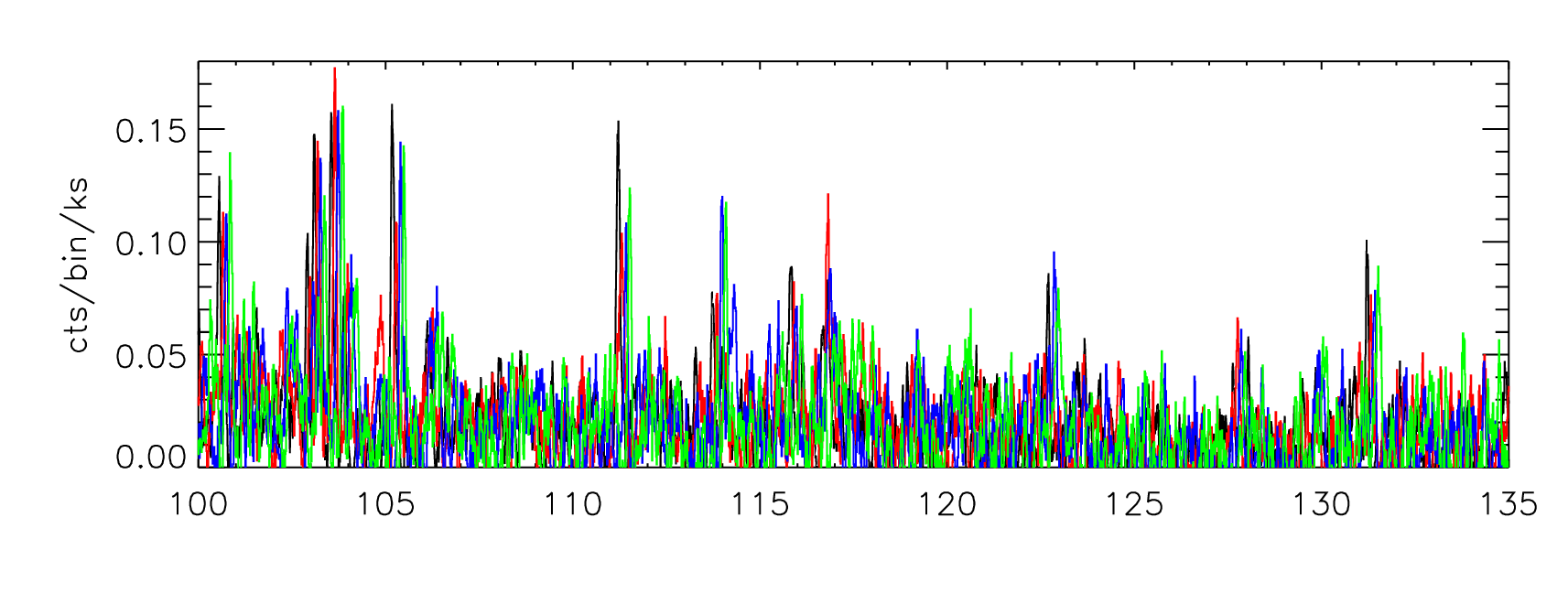}\vspace{-1.1cm}}
\centerline{\includegraphics[scale=0.55,angle=90]{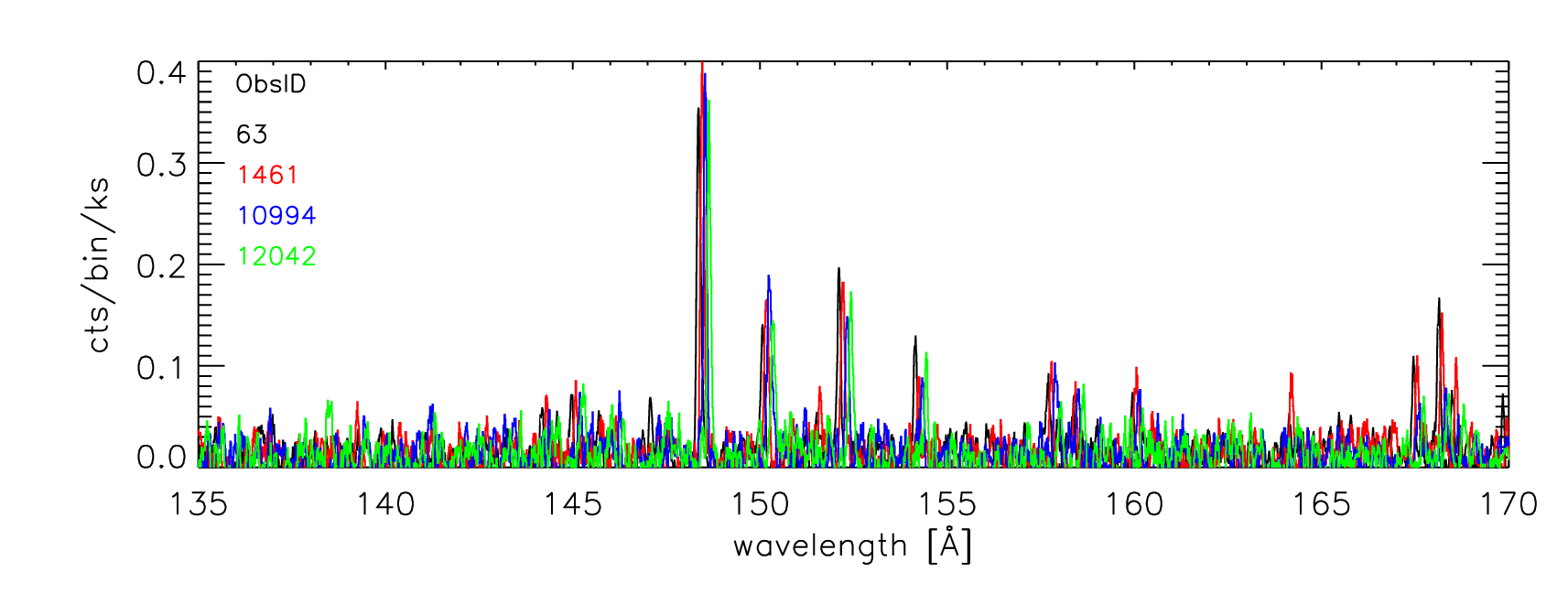}\vspace{-0.5cm}}
\caption{Comparison of \cha\ LETGS spectra of Procyon obtained in four different
  observations (a different color is used for each observation, as
  labeled in the bottom plot). 
  For easier comparison each spectrum is shifted by +0.1\AA\ with
  respect to the preceding one.  \label{fig:Procyon_4spec}}
\end{figure*}

The details of the four \cha\ observations are listed in
Table~\ref{tab:obs}.  Light curves derived from the 0th order photon events, 
binned at an interval of 1ks, are illustrated in Figure~\ref{fig:lc} and the 
spectra extracted from each epoch are shown in
Figure~\ref{fig:Procyon_4spec}.  Both light curves and spectra show  the
X-ray emission  of Procyon to be remarkably constant, both in terms
of flux level and spectral characteristics, and on all observed time
scales, from 1ks  to much longer timescales of over 10 years.
Close inspection of the spectra does reveal some differences, 
especially between 49 and 69\AA, and above 160\AA.  These can be 
ascribed to the different effective areas at those wavelengths, due to 
the wavelengths in the spectra at which the HRC-S plate gaps occur.  
These vary from epoch to epoch because of the secular aim point drift of 
\cha\ relative to the detector coordinate system.
We found no evidence for any significant differences between the spectra 
taken at different epochs outside of the wavelength regions affected by plate gaps.
This lack of variability over timescales ranging from $\sim 1$ks to
about 10 years is remarkable for the X-ray emission of late-type
stars.  Although the sampling is sparse, this constancy hints at a lack of
a large amplitude magnetic activity cycle analogous to that observed for 
the Sun and seen in X-rays in other low and moderately active stars 
\citep[e.g.,][]{Hempelmann06,Favata08,Ayres09}.

\section{Analysis Methods and Results}
\label{ss:results}

In order to assess completeness and accuracy of the CHIANTI atomic
database we proceeded as follows: (1) we selected a set of lines
formed over a wide temperature range, unblended, and with reliable
atomic data; (2) we reconstructed the emission measure distribution,
EM(T), of the emitting coronal plasma using the measured fluxes, and
finally (3) we synthesized the model spectrum using CHIANTI and the
EM(T) derived from the data and compared it with the observed spectrum
at all wavelengths. 

The line fluxes used for determining the thermal distribution of the
plasma are listed in Table~\ref{tab:lineflx}. The selected lines are
formed over a broad temperature range, as indicated by their
temperature of maximum formation, covering roughly the interval from
$\log (T[K]) \sim 5.7$ to $\log (T[K]) \sim 6.9$  (see
Table~\ref{tab:lineflx}).  

\begin{deluxetable}{rlccc} 
\footnotesize
\tablecolumns{5} 
\tablewidth{0pc} 
\tablecaption{Identification and measured fluxes from {\em Chandra} LETGS spectral 
	observations of Procyon ($\sim 275$~ks).
	\label{tab:lineflx}} 
\tablehead{ 
\colhead{\ll}  & \colhead{Ion} & \colhead{$\log (T_{\rm max} [{\rm K}])$} &  
	\colhead{flux}  &  \colhead{flx$_{\rm
            pred}$/flx$_{\rm obs}$}  \\ 
\colhead{[\AA]}  & \colhead{} & \colhead{} &  
	\colhead{[$10^{-6}$ \flxu]}  &  \colhead{}  
}
\startdata 
  
12.1324  &  \nex    &  6.75  &  $5.82^{u} \pm 1.8$  &  0.56 \\ 
13.4417  &  \neix   &  6.60  &  $18.6 \pm 1.8$  &  1.2 \\ 
13.6852  &  \neix   &  6.60  &  $14.4 \pm 1.4$  &  1.2 \\ 
15.0070  &  \fexvii  &  6.80  &  $24.2 \pm 1.8$  &  1.1 \\ 
15.9958  &  \oviii   &  6.50  &  $17.6 \pm 1.8$  &  1.2 \\ 
16.0590  &  \fexviii  &  6.90  &  $2.15^{u} \pm 1.3$ &  0.62 \\ 
16.7783  &  \fexvii  &  6.70  &  $16.0 \pm 1.5$  &  1.2 \\ 
17.0768  &  \fexvii  &  6.70  &  $53.6 \pm 2.9$  &  0.82 \\ 
18.6271  &  \ovii   &  6.30  &  $46.7 \pm 2.8$  &  0.85 \\ 
18.9651  &  \oviii  &  6.50  &  $224 \pm 10$  &  0.95 \\ 
21.6077  &  \ovii   &  6.30  &  $379 \pm 15$  &  1.1 \\ 
21.8064  &  \ovii   &  6.30  &  $95.0 \pm 4.9$  &  1.0 \\ 
22.0979  &  \ovii   &  6.30  &  $335 \pm 11$  &  0.87 \\ 
24.7879  &  \nvii   &  6.30  &  $102 \pm 6.2$  &  0.84 \\ 
28.4630  &  \cvi    &  6.15  &  $55.9 \pm 4.0$  &  0.87 \\ 
28.7769  &  \nvi    &  6.15  &  $91.5 \pm 5.1$  &  0.86 \\ 
29.0736  &  \nvi    &  6.10  &  $30.9 \pm 3.0$  &  0.83 \\ 
29.5386  &  \nvi    &  6.10  &  $51.7 \pm 4.1$  &  0.96 \\ 
30.4453  &  \caxi  &  6.30  &  $34.5 \pm 3.7$  &  1.3 \\ 
33.7451  &  \cvi    &  6.10  &  $508 \pm 19$  &  1.1 \\ 
34.9869  &  \cv     &  6.00  &  $23.2 \pm 2.9$  &  0.86 \\ 
39.2737  &  \sxi    &  6.30  &  $69.5 \pm 5.8$  &  0.75 \\ 
40.7251  &  \cv     &  5.95  &  $190 \pm 16$  &  0.55 \\ 
43.7323  &  \sixi   &  6.30  &  $68.4 \pm 3.5$  &  0.90 \\ 
43.9891  &  \sixii   &  6.30  &  $42.2 \pm 3.1$  &  0.81 \\ 
44.1380  &  \sixii   &  6.30  &  $61.6 \pm 3.7$  &  1.1 \\ 
45.4889  &  \sixii   &  6.30  &  $18.8 \pm 1.9$  &  0.96 \\ 
45.6569  &  \sixii   &  6.30  &  $40.6 \pm 2.7$  &  0.91 \\ 
46.3715  &  \s9      &  6.20  &  $31.9 \pm 2.7$  &  1.5 \\ 
47.2280  &  \s9      &  6.10  &  $32.7 \pm 2.9$  &  1.3 \\ 
47.6279  &  \six     &  6.20  &  $48.0 \pm 3.7$  &  1.0 \\ 
47.7801  &  \sx      &  6.20  &  $25.4 \pm 2.7$  &  0.78 \\ 
49.1922  &  \sixi    &  6.25  &  $125 \pm 10$  &  1.1 \\ 
50.5078  &  \six     &  6.20  &  $166 \pm 10$  &  0.77 \\ 
61.0172  &  \siviii  &  6.00  &  $154 \pm 11$  &  0.57 \\ 
63.1306  &  \six     &  6.10  &  $70.1 \pm 6.6$  &  0.96 \\ 
63.2662  &  \six     &  6.10  &  $126 \pm 12$  &  0.95 \\ 
65.6395  &  \six     &  6.10  &  $41.9 \pm 4.1$  &  0.87 \\ 
65.8299  &  \six     &  6.10  &  $67.0 \pm 5.0$  &  1.1 \\ 
67.2169  &  \siix     &  6.00  &  $57.0 \pm 5.0$  &  1.0 \\ 
71.9973  &  \svii     &  5.80  &  $51.9 \pm 4.6$  &   1.0 \\ 
72.2828  &  \mgix  &  6.00  &  $134 \pm 9.7$  &  1.1 \\ 
72.8687  &  \svii     &  5.80  &  $43.4 \pm 4.5$  &  0.86 \\ 
77.7143  &  \mgix  &  6.00  &  $106 \pm 7.0$  &  1.1 \\ 
88.0453  &  \neviii  &  5.85  &  $173 \pm 10$  &  0.71 \\ 
96.0970  &  \fex     &  6.10  &  $139 \pm 10.4$  &  0.94 \\ 
97.4721  &  \nevii   &  5.80  &  $62.3 \pm 6.7$  &  0.68 \\ 
98.0717  &  \neviii  &  5.80  &  $145 \pm 10$  &  0.89 \\ 
98.2259  &  \neviii  &  5.80  &  $238 \pm 16$  &  1.1 \\ 
97.1081  &  \fex     &  6.10  &  $58.6 \pm 6.9$  &  0.61 \\ 
102.898  &  \neviii  &  5.80  &  $84.3 \pm 8.1$  &  0.85 \\ 
103.087  &  \neviii  &  5.80  &  $137 \pm 10$  &  1.1 \\ 
103.547  &  \feix    &  5.90  &  $208 \pm 13$  &  0.78 \\ 
105.183  &  \feix    &  5.90  &  $157 \pm 11$  &  0.59 \\ 
127.649  &  \nevii   &  5.75  &  $113 \pm 12$  &  0.65 \\ 
130.940  &  \feviii   &  5.70  &  $69.5 \pm 12$  &  1.4 \\ 
131.227  &  \feviii   &  5.70  &  $172 \pm 15$  &  1.1 \\ 
148.358  &  \nixi    &  6.10  &  $908 \pm 37$  &  1.2 \\ 
152.117  &  \nixii   &  6.20  &  $420 \pm 26$  &  0.86 \\ 
154.143  &  \nixii   &  6.20  &  $276 \pm 21$  &  0.67 \\ 
157.684  &  \nixiii  &  6.30  &  $156 \pm 18$  &  1.7 \\ 
167.459  &  \feviii  &  5.70  &  $659 \pm 59$  &  1.1 \\ 
168.130  &  \feviii  &  5.70  &  $971 \pm 72$  &  1.2 \\ 
168.497  &  \feviii  &  5.70  &  $583 \pm 57$  &  1.0 \\ 
168.895  &  \feviii  &  5.70  &  $317 \pm 44$  &  0.98 \\ 

\enddata 
\tablenotetext{u}{The label ``u'' associated with a flux value indicates an upper limit. }
\end{deluxetable}

\begin{figure*}[!ht]
\centerline{\includegraphics[scale=0.7,angle=90]{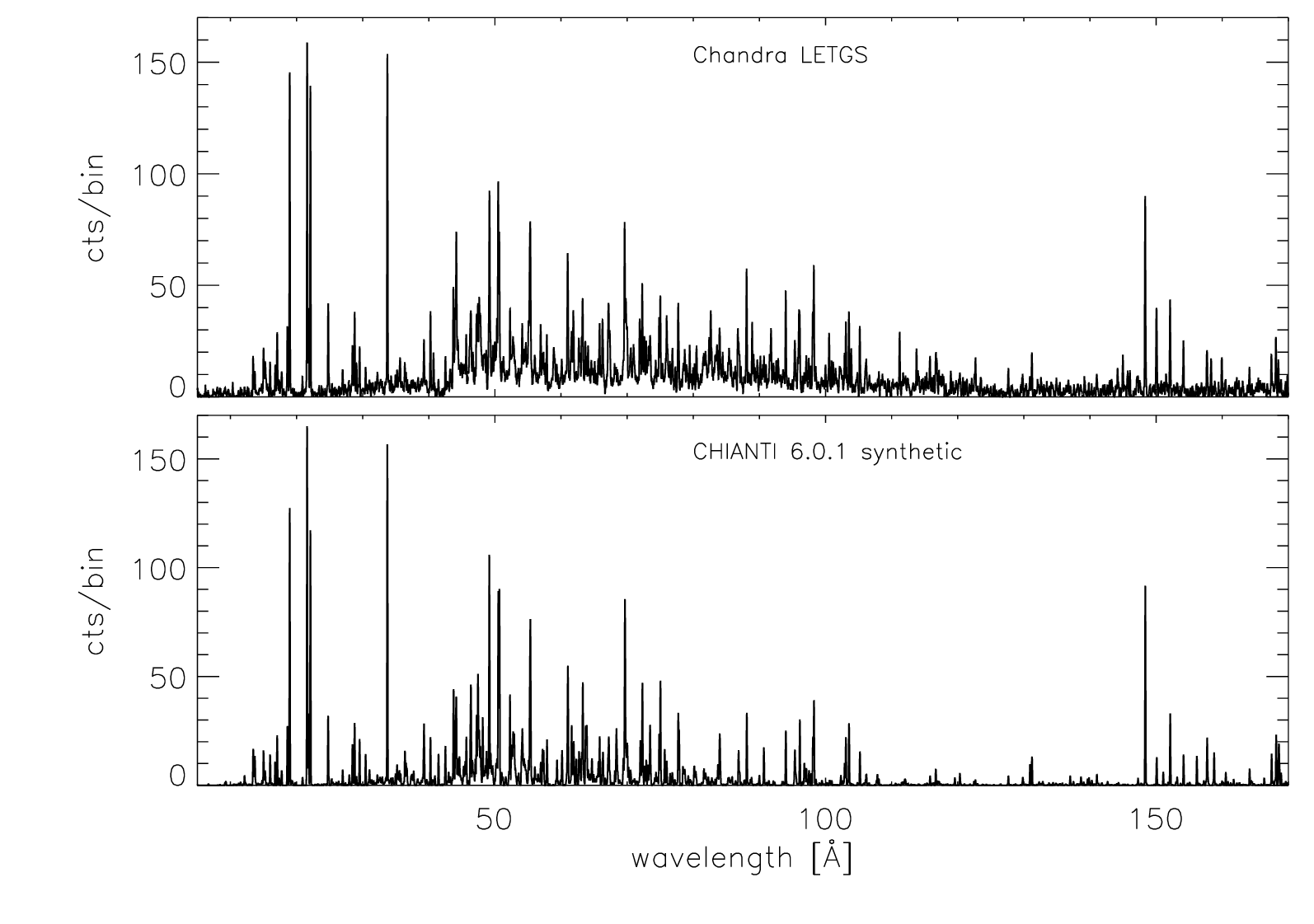}\vspace{-0.3cm}}
\caption{{\em Top:} \cha\ LETGS spectrum of Procyon, obtained with a total exposure
  time of approximately 275~ks (Observation ID: 63, 1461, 10994, 12042). {\em
    Bottom:} Model spectrum synthesized using CHIANTI 6.0.1 and the
  emission measure distribution derived from the measured line fluxes
  listed in Table~\ref{tab:lineflx}, and plotted in
  Figure~\ref{fig:emt}.  An expanded version of the model and observed
  spectra, allowing a more detailed comparison, is presented in
  Figure~\ref{fig:spec_mod}. 
	\label{fig:spectrum}} 
\end{figure*}

We determined the emission measure distribution by using an iterative
method based on a Markov-chain Monte Carlo (MCMC) algorithm (see
\citealt{Kashyap98} for details on assumptions and approximations, and
also \citealt{Testa11} for further description of the characteristics of
the method). 
The emission measure distribution is reconstructed using the line
emissitivies from CHIANTI v.6.0.1 \citep{chianti,chianti6}, and
assuming the ionization balance of \cite{Bryans09}.

Previous analyses of X-ray and EUV spectra of Procyon have indicated
that the element abundances of its coronal plasma are close to
its photospheric abundances \citep[e.g.,][]{Drake95,Raassen02,Sanz04},
which in turn are similar to solar photospheric abundances
\citep[e.g.,][]{Steffen85,Drake.Laming95b,Bruntt10}. 
By using an abundance diagnostic technique based on temperature
insensitive line ratios \citep{DrakeTesta05,Huenemoerder09} we derive
an estimate of the Ne/O abundance ratio of $1.3 \times$ the solar
photospheric value of \citet{GrevesseSauval}.  This diagnostic technique
uses the measured fluxes of the strong H-like and He-like transitions,
lying at the short wavelength range of the \cha\
spectra.  Procyon is characterized by a rather inactive
and cool corona, compared with the more active stellar coronae usually
observed with \cha. Therefore, even with the long accumulated exposure time of the
spectra analyzed here ($\sim 275$~ks) the H-like and
He-like lines of the higher Z elements (mainly Mg, Si, S, Fe), which
are formed at temperatures higher than the corresponding transitions
of Ne and O, are not detected.

\begin{figure}[!ht]
\centerline{\includegraphics[scale=0.55]{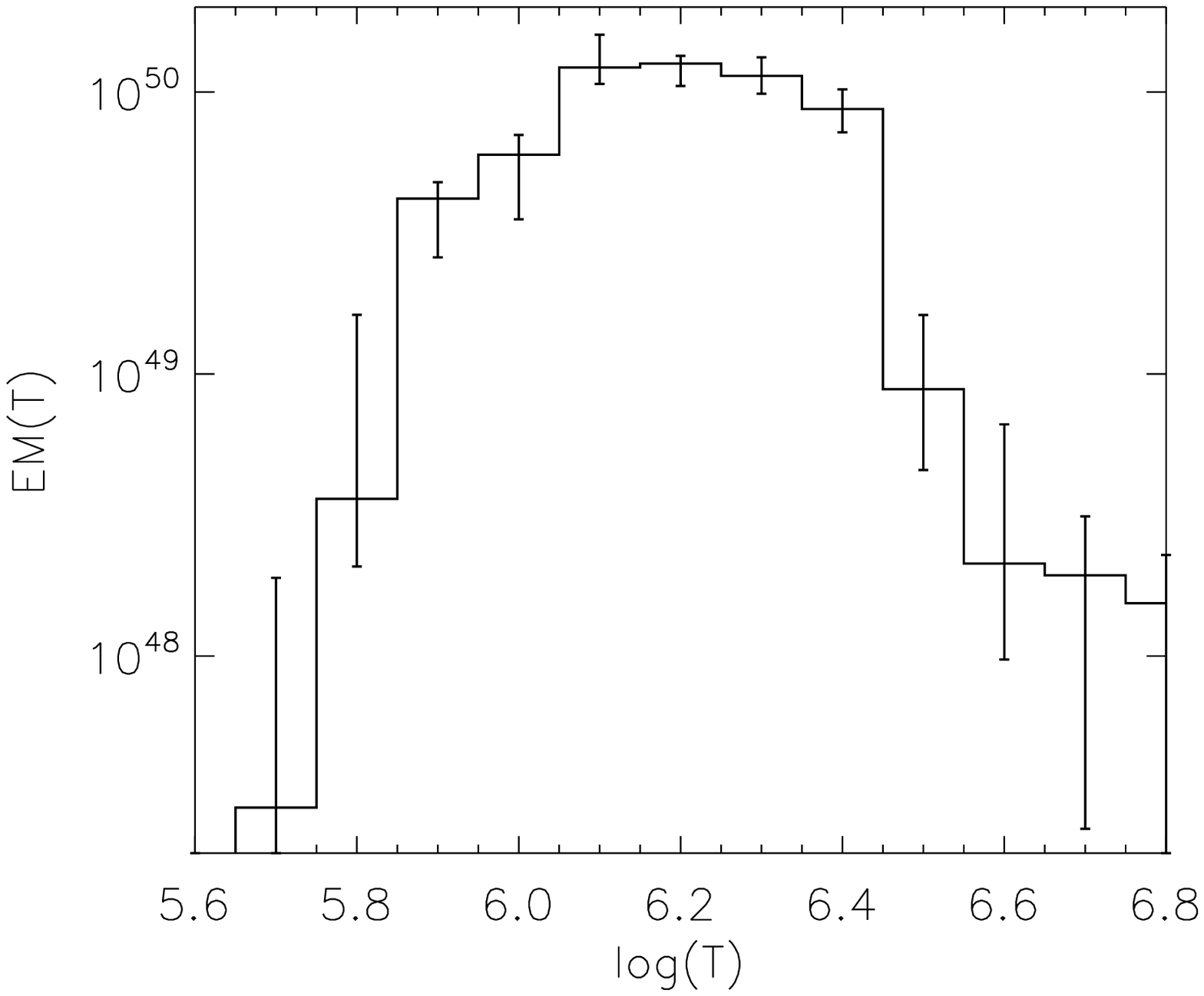}\vspace{-0.3cm}}
\caption{Emission measure distribution, EM(T), derived applying a
  Markov Chain Monte Carlo (MCMC) iterative method to the measured
  fluxes of the lines listed in Table~\ref{tab:lineflx}.  
	\label{fig:emt}}
\end{figure}

We ran the procedure to derive the emission measure distribution
several times and 'manually' adjusted the abundances of those elements
if the agreement with the model was unsatisfactory and the predicted
line fluxes for a given element were consistently lower or higher than the
observed fluxes.
The model we deemed to be best is characterized by
photospheric abundances \citep{GrevesseSauval}, except for  N and Ne
which are enhanced by 30\% and Mg, Ca, and Fe that are enhanced by 50\%.
While perhaps slightly higher than current assessments of solar-like 
photospheric abundances in Procyon, our values are not significantly 
different considering the combined systematic and random errors of 
photospheric and coronal analyses. 
We express our derived abundances relative to the reference set 
of abundances of \citet{GrevesseSauval}, but the choice of reference
is arbitrary.  We adopted the set of \citet{GrevesseSauval} for 
easier comparison with other works in the literature.

\begin{figure*}[!ht]
\centerline{\includegraphics[scale=0.65,angle=90]{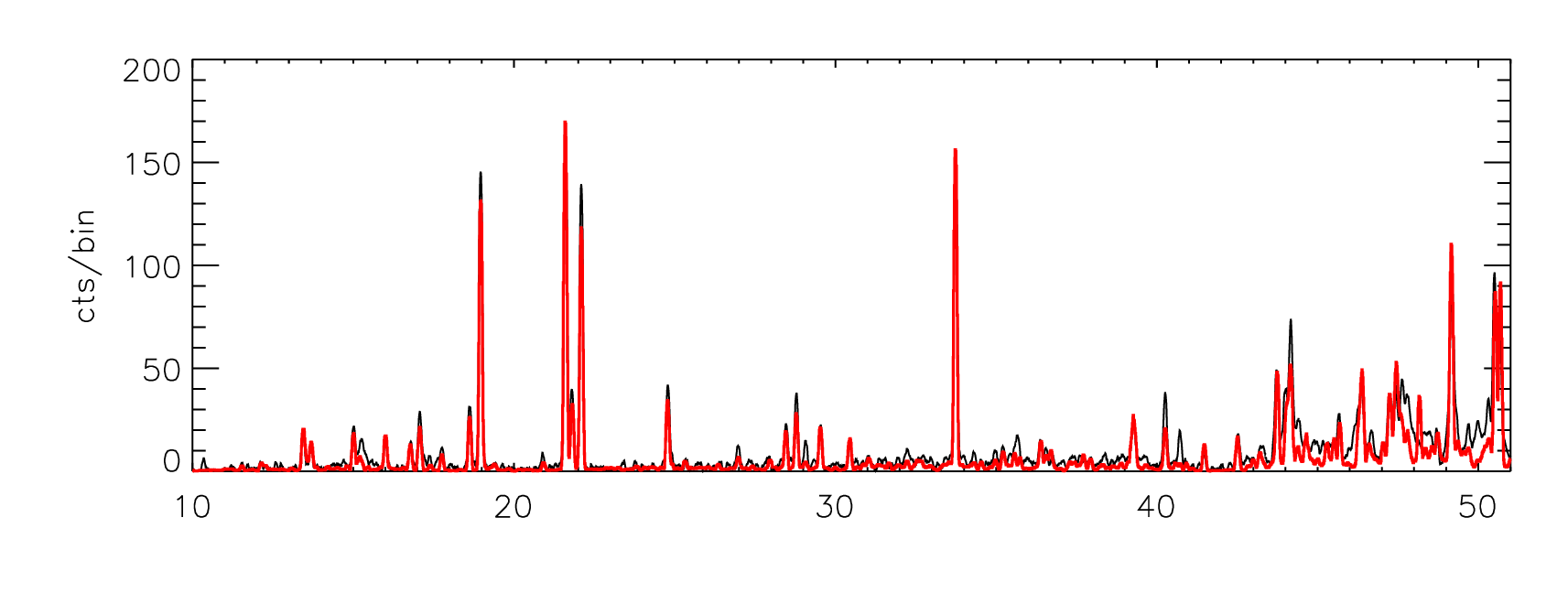}\vspace{-1.3cm}}
\centerline{\includegraphics[scale=0.65,angle=90]{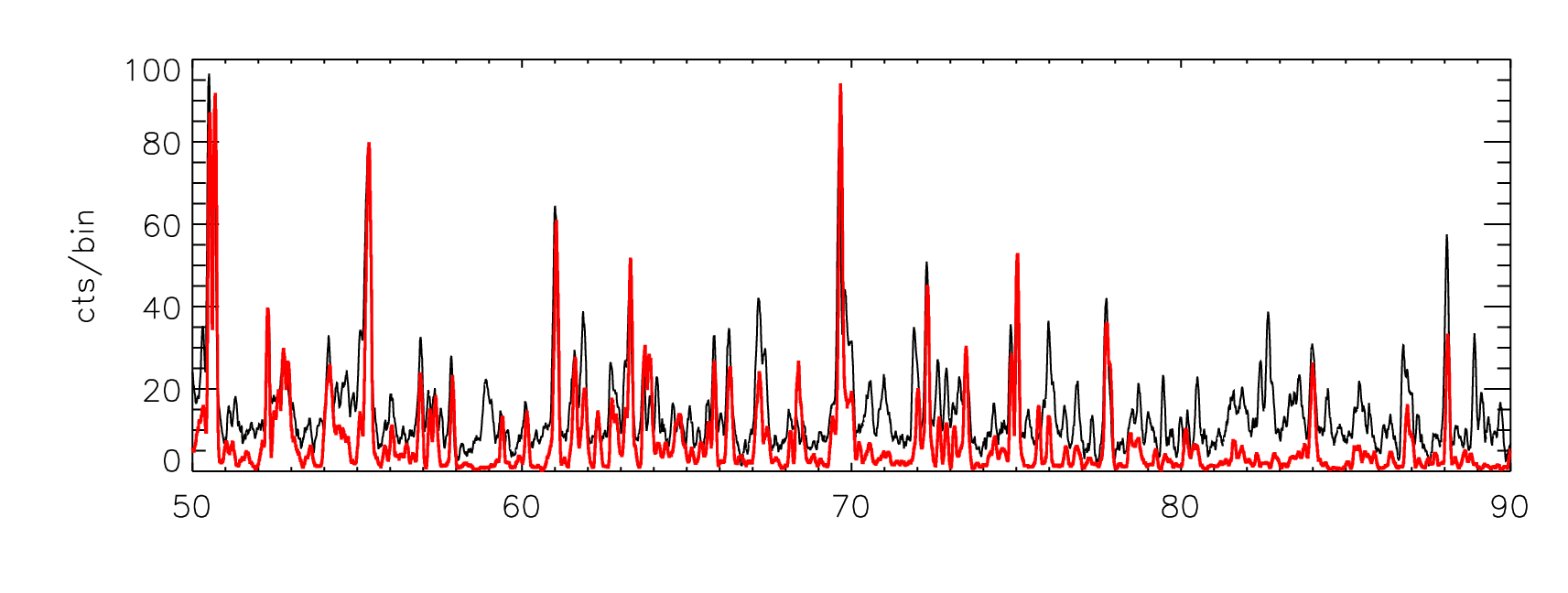}\vspace{-1.3cm}}
\centerline{\includegraphics[scale=0.65,angle=90]{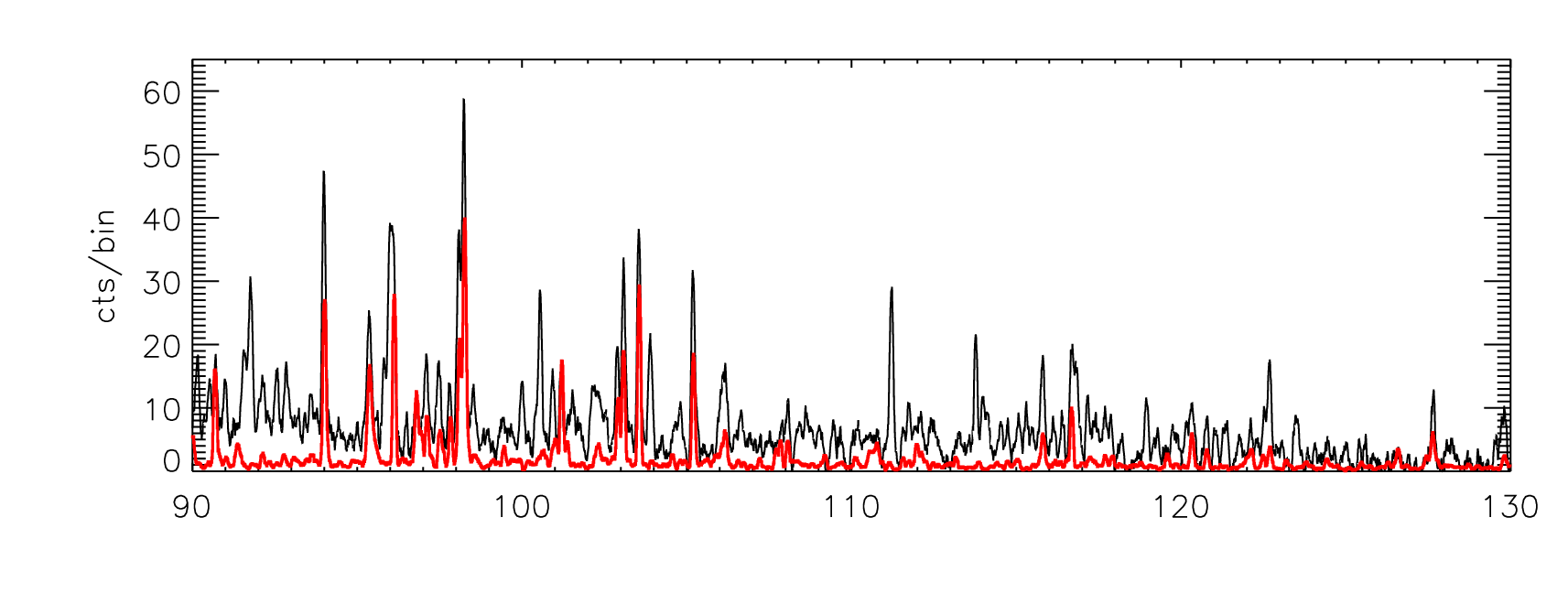}\vspace{-1.3cm}}
\centerline{\includegraphics[scale=0.65,angle=90]{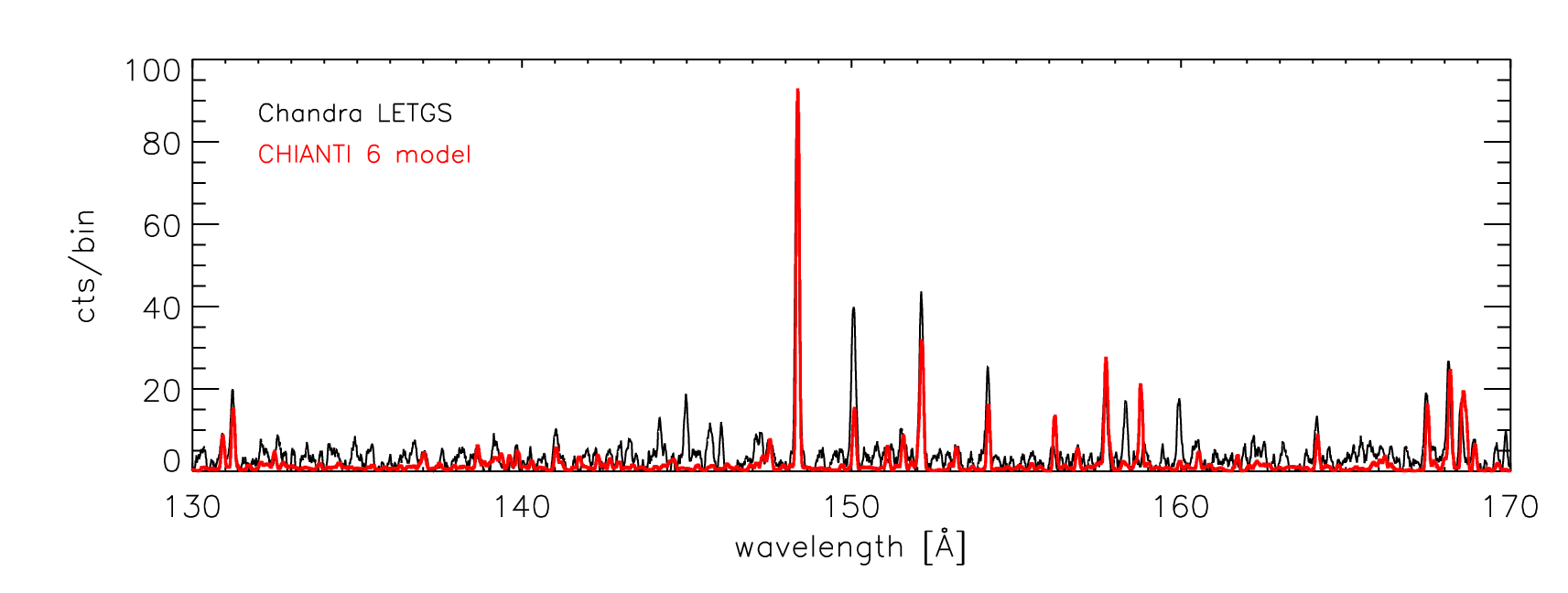}\vspace{-0.1cm}}
\caption{Comparison of \cha\ LETGS observations (black) and CHIANTI 6
  model (red; including orders $\pm$ 1,2,3) synthesized using the EM(T) shown in
  Figure~\ref{fig:emt}, derived using the line fluxes of Table~\ref{tab:lineflx}. 
	\label{fig:spec_mod}}
\end{figure*}

The emission measure distribution derived from the measured fluxes is 
illustrated in Figure~\ref{fig:emt}.  It is a rather smooth and broad
function of temperature, with most material found around $\log
(T[K]) \sim 6.1-6.3$, i.e.\ at temperatures typical of non-flaring
solar coronal plasmas. This emission measure distribution we derive 
from the long \cha-LETGS exposure compares well with those derived
for the same source by \citet{Drake95} from {\em EUVE} observations, and by
\citet{Raassen02} from the first two LETGS observations and 
\xmm\ spectra, having a similar peak temperature ($\sim 1.5$MK),
emission measure value ($\sim 10^{50}$~cm$^{-3}$), and width of the 
distribution.
Some small differences are present either side of the EM distribution
peak at temperatures of $\log T = 6.0-6.4$.  Such differences might be 
expected based on the different global model fitting approach adopted by 
Raassen et al, in addition to the more up-to-date ionization balance and 
collisional excitation data used in the current study. 
In order to evaluate how well the EM(T) of Figure~\ref{fig:emt}
reproduces the observed emission,  the measured fluxes
are compared with the fluxes predicted by the model EM(T) in 
Table~\ref{tab:lineflx}.  Agreement is  satisfactory and typically within $\sim 30$\%, 
which is in accordance with the assessment by \citet{Drake95} of typical atomic 
data uncertainties. 

\begin{figure}[!ht]
\centerline{\includegraphics[scale=0.55]{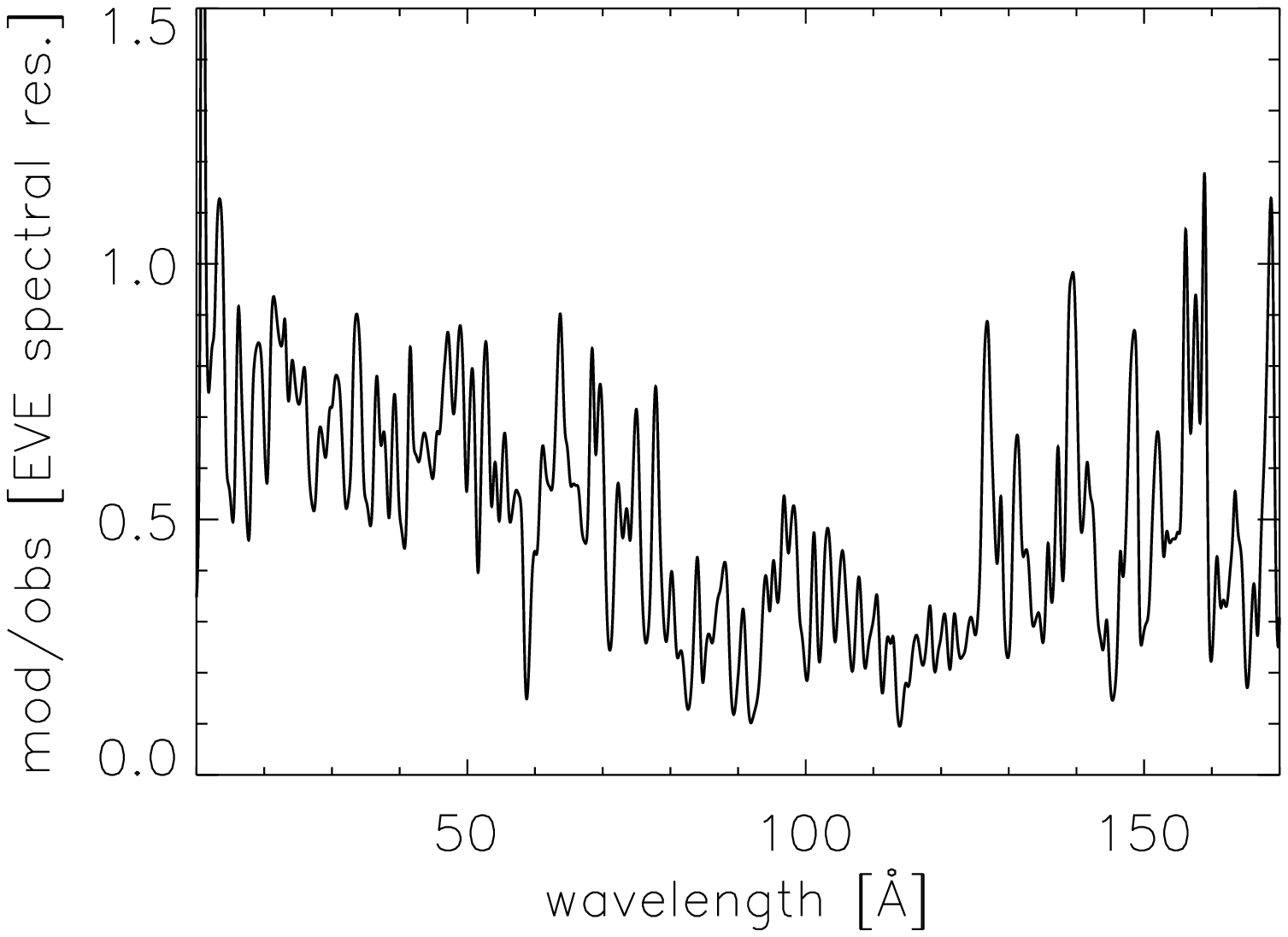}
  \vspace{-0.2cm}}
\caption{Ratio of the CHIANTI model spectrum to the observed spectrum,
both degraded to the EVE spectral resolution. \label{fig:comp_eve_res}}
\end{figure}

In Figure~\ref{fig:spectrum} (bottom panel) we show the model spectrum
synthesized from the EM(T) of Figure~\ref{fig:emt} using the
CHIANTI database (v.6.0.1) and the element abundances and
ionization equilibrium described above. 
In order to allow a more detailed comparison of the model with the
observed spectrum, in Figure~\ref{fig:spec_mod} we show the two
spectra superimposed (observed spectrum in black and model in red).

\begin{figure*}[!ht]
\centerline{\includegraphics[scale=0.5]{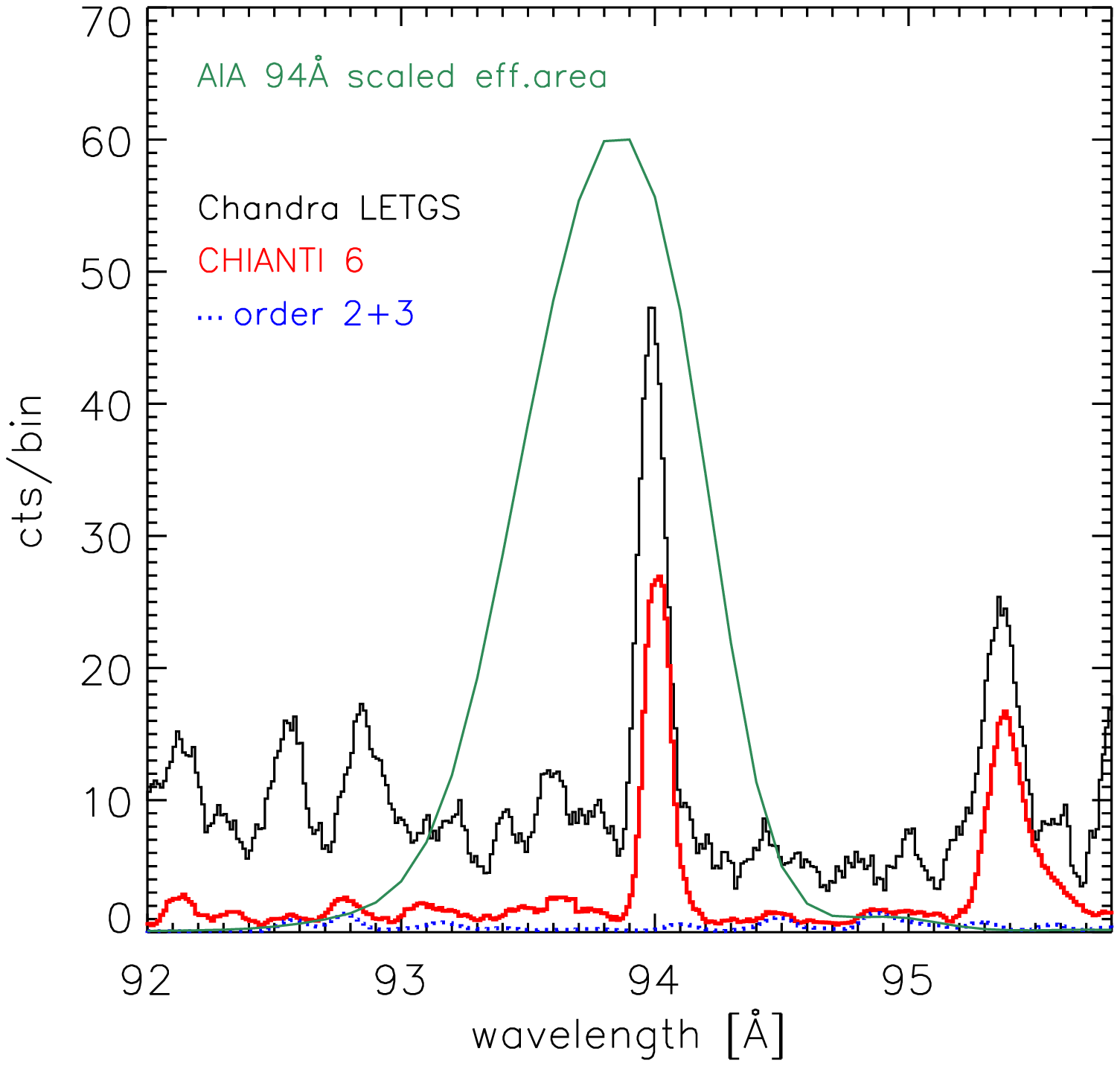}
  \includegraphics[scale=0.5]{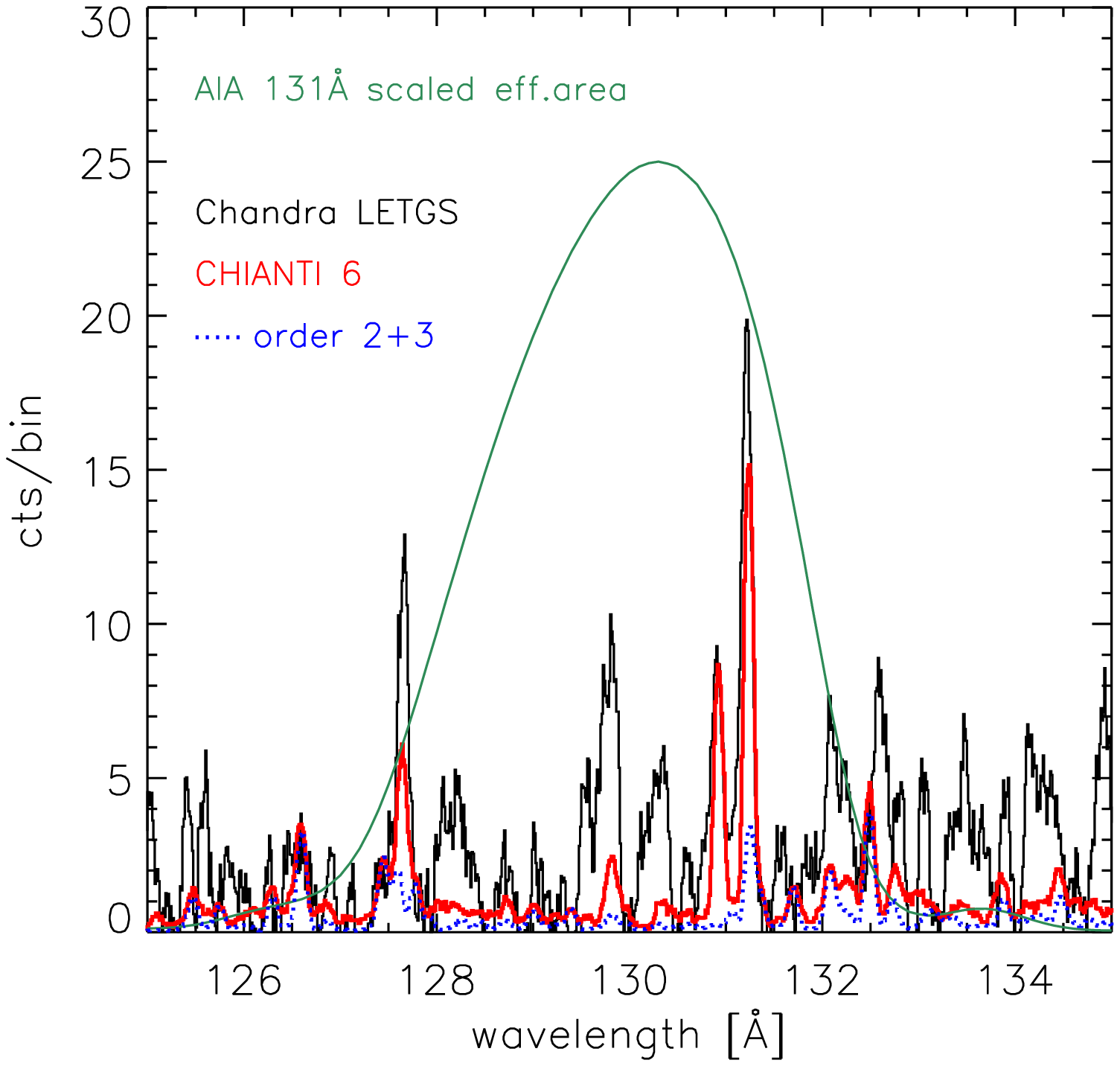}
  \vspace{-0.2cm}}
\centerline{\includegraphics[scale=0.5]{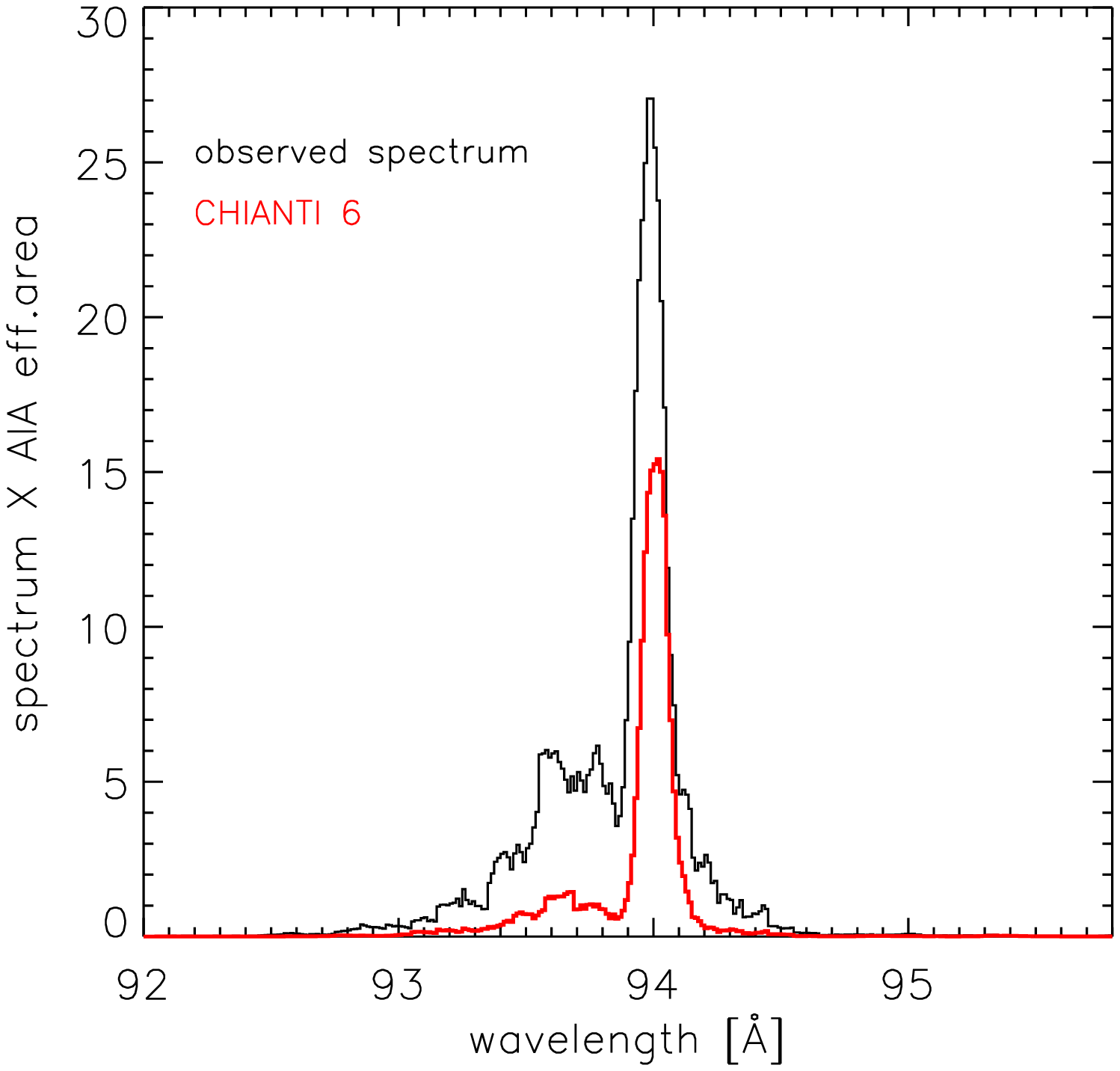}
  \includegraphics[scale=0.5]{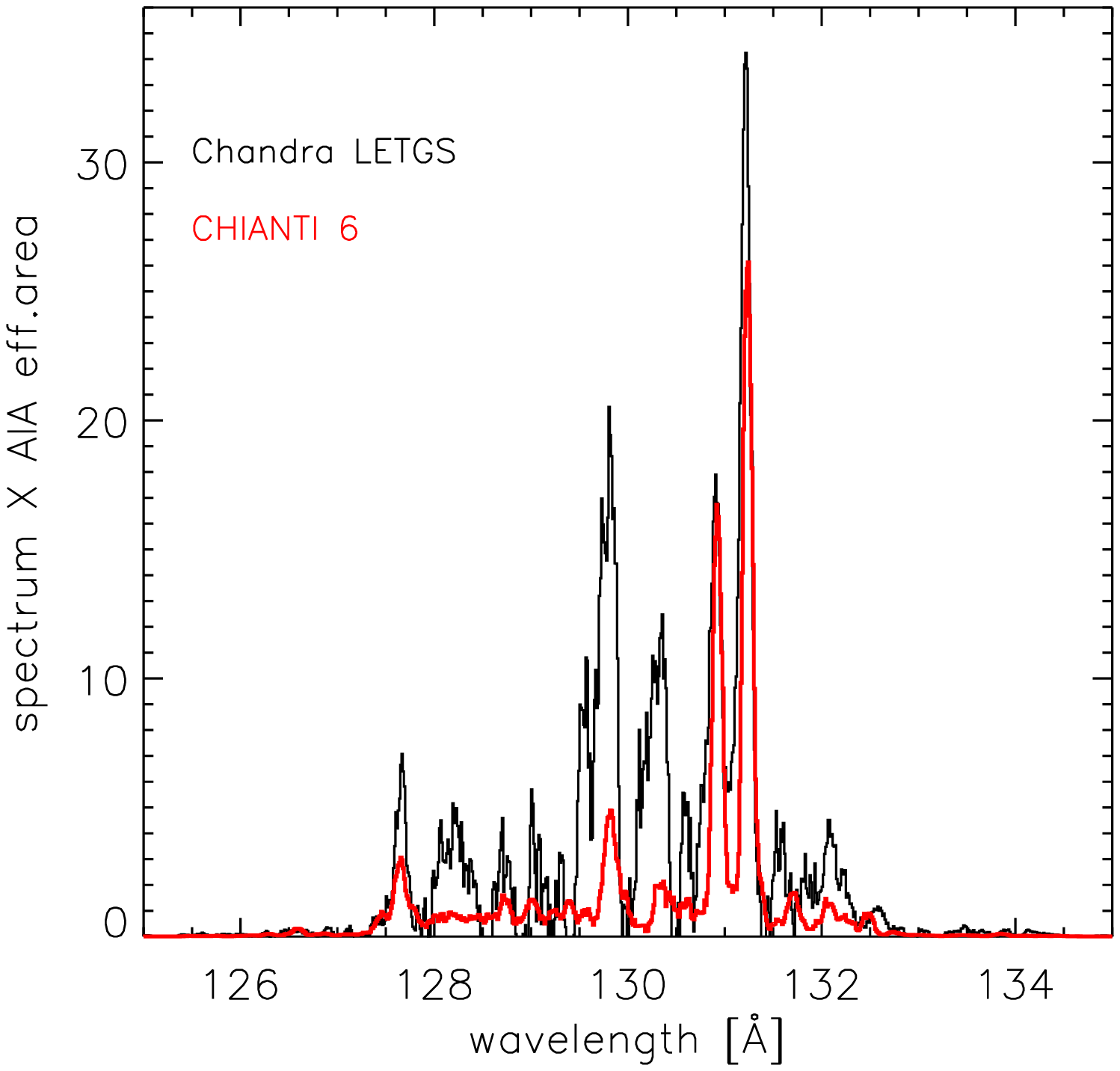}
\vspace{-0.1cm}}
\caption{{\em Top}: observed LETGS spectrum of Procyon (black curve) and 
  model (total of orders 1, 2, and 3, positive and negative; red curve) in 
  the wavelength regions observed by the shortest wavelength narrow-bands
  of the Atmospheric Imaging Assembly (AIA) on the Solar Dynamics
  Observatory. Scaled effective areas for the 94\AA, and 131\AA, AIA
  passbands (green curves) are superimposed to indicate the sensitivity of 
  AIA to the spectral features in each wavelength range.  We also plot the 
  high order ($\pm 2,3$) contribution to the spectrum model.
  {\em Bottom:} Observed and model spectrum (black and red respectively)
  multiplied by the AIA effective area in the 94\AA\ channel ({\em left}) and
  131\AA\ channel ({\em right}). These curves show the contribution of the 
  different spectral features to the AIA observed intensity.
	\label{fig:spec_mod_aia}}
\end{figure*}

Figure~\ref{fig:spectrum} and \ref{fig:spec_mod} demonstrate 
quite a favorable comparison at the shorter wavelengths. At longer wavelengths, 
especially in the $\sim 80-110$\AA\ range, it is apparent that the observed 
spectrum contains a large number of lines, strong and weak, missing in the 
CHIANTI spectral model. 
This missing flux was first noted in earlier work on {\em EUVE} stellar spectra,
including that of Procyon (see, e.g., discussions by 
\citealt{Drake96,Drake.Kashyap01}) and was also touched upon by 
\citet{Raassen02} in their analysis of the first two LETGS Procyon observations.
We also note that analyses of \sdo-EVE spectra appear to show the
same effect: spectral models are in good agreement with the
observed spectra at EUV wavelengths ($\gtrsim 150$\AA) whereas they
represent a poor match to the observed emission in the $\sim
50-150$\AA\ range (H.Warren, private communication).  These findings
strongly suggest that the missing flux is not due to inadequate
modeling of transitions included in current databases but to incompleteness 
atomic databases at those wavelengths. 
In order to estimate the effect of the flux apparently lacking in the 
model spectrum on the analysis of EVE spectra, we degraded both 
the model and the observed spectrum to EVE spectral resolution and 
plot their ratio in Figure~\ref{fig:comp_eve_res}.  This plot shows 
that the model underestimates the observed flux by a variable factor
ranging from $\approx 1.5$, at short wavelengths below $\sim$50\AA,
up to $\approx$5-7 in the $\sim 70-125$\AA\ range.
We defer a more detailed discussion to Drake et al. (2011, in preparation).

\begin{figure}[!ht]
\centerline{\includegraphics[scale=0.5]{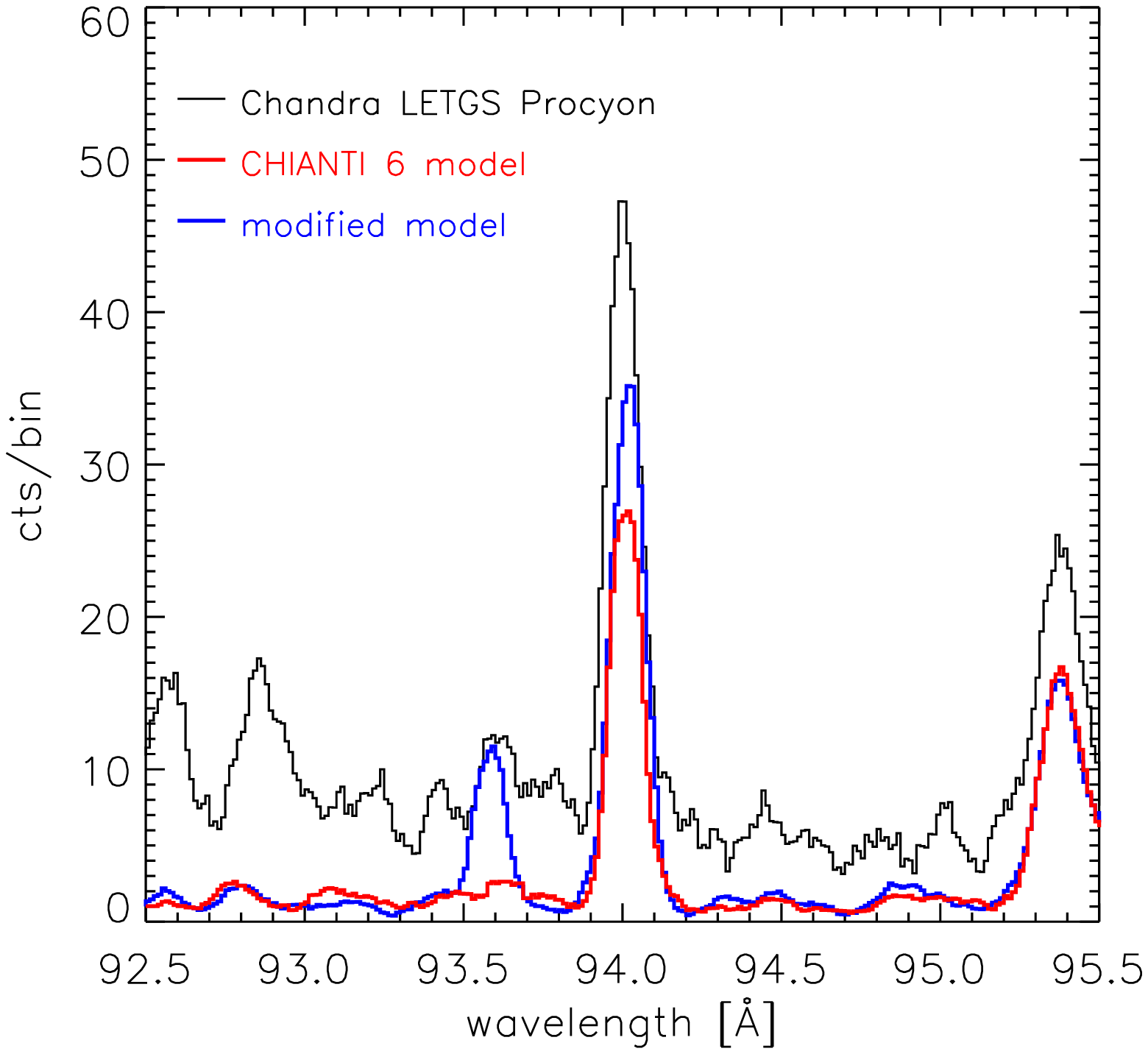}
\vspace{-0.3cm}}
\caption{A modified spectral model (blue), with the addition of an empirical
  correction to the CHIANTI model including \feix\ transitions
  observed in laboratory experiments (see text and Foster \& Testa 2011) 
  and not present in the CHIANTI atomic database, is superimposed to the 
  CHIANTI 6 model (red) and observed spectra (black).
	\label{fig:spec_mod_emp}}
\end{figure}

Figure~\ref{fig:spec_mod_aia} shows the comparison of the spectral model
and observations in the wavelength ranges of the AIA passbands centered on 
94\AA\
and 131\AA\ (171\AA\ is at the edge of the sensitivity of \cha-LETGS).
These plots demonstrate that the model lacks a significant portion
of the observed flux, especially in the 94\AA\ range. In particular,
in the 94\AA\ band, not only is the strongest feature, \fexviii\ 93.93\AA +
\fex\ 94.01\AA, 
significantly underestimated by the model, but the observed level of
emission is consistently higher than the model throughout the bandpass.  
This is likely due to the
contribution of a large number of relatively weak lines 
that form a pseudo-continuum which becomes comparable in flux 
to the resolved lines. 
In the 131\AA\ band, the disagreement is less severe: the predicted
fluxes of the strong \feviii\ lines are close to the measured
values, though the flux contribution of weaker lines is underpredicted (or
completely lacking) in the model.
We note that neither the uncertainties on the
background subtraction nor the possible contamination from the 
overlapping higher spectral orders can explain the observed discrepancies (in 
Figure~\ref{fig:spec_mod_aia} we also show separately the 
modeled high order contribution, as a blue dotted curve).

In order to estimate what the effect of the missing flux would be on
an AIA observation of coronal plasma with Procyon-like X-ray-EUV
emission, we convolved the observed and modeled spectrum with the 
responses of the AIA 94\AA\ and 131\AA\ narrow passbands as a function of 
wavelength.  We find that the AIA flux for a Procyon-like source would be
underestimated by the spectral model by a factor of roughly 3 in the
94\AA\ passband and 1.9 in the 131\AA\ passband.
The plots of the observed and model spectra folded with the AIA responses in 
the 94\AA\ and 131\AA\ channels (bottom panels of Figure~\ref{fig:spec_mod_aia}), 
clearly show which spectral features are more relevant, contributing more 
significantly to the AIA intensity.

Laboratory experiments with the Lawrence Livermore electron beam ion
trap, EBIT, \citep{Beiersdorfer99,Lepson02} focusing on the wavelength
range $\approx 60-140$\AA\ investigated the presence of Fe lines
(Fe\,{\sc vii-x}) possibly not included in current atomic databases and
potentially important for the interpretation and modeling of solar and
stellar soft X-ray-EUV spectra. We reviewed their findings to explore
whether any of their detected emission lines that are still absent from 
atomic databases might affect significantly the 94\AA\ and 131\AA\
AIA wavelength ranges. 
In their list of emission lines, we find that there are two
transitions with potentially significant impact for the 94\AA\ AIA
passband. These are two \feix\ lines -- $3p^55f \rightarrow 3p^53d$ 
transitions -- with measured wavelengths of 
$\lambda$ 93.59\AA\ and 94.07\AA\ respectively \citep{Lepson02},
and intensities of roughly 0.25 and $0.3 \times$ the intensity of the 
\feix\ emission at $103.55$\AA.  We note that the intensities reported 
in these laboratory experiments cannot be directly translated into relative
intensity expected for the emission of coronal plasma, as in EBIT
experiments the plasma conditions can significantly deviate from the
conditions generally assumed for coronal thermal plasma (for instance
the electrons have non-Maxwellian distribution). 
\cite{Foster11} used the Flexible Atomic Code to carry out 
calculations for these \feix\ transitions, and we explore the possible 
relevance of these lines to explain the missing flux, by recomputing 
the model adding these additional \feix\ contributions in the 94\AA\ range 
to the synthetic CHIANTI model spectrum. This new spectral model including the
\feix\ transitions, shown in Figure~\ref{fig:spec_mod_emp} (blue
line), is in much better agreement with the observed spectrum in that
narrow wavelength range, and reduces the flux discrepancy from the factor 
$\sim 3$ for the pure CHIANTI 6 spectrum to roughly
$80$\%.

\begin{figure}[!ht]
\centerline{\includegraphics[scale=0.45]{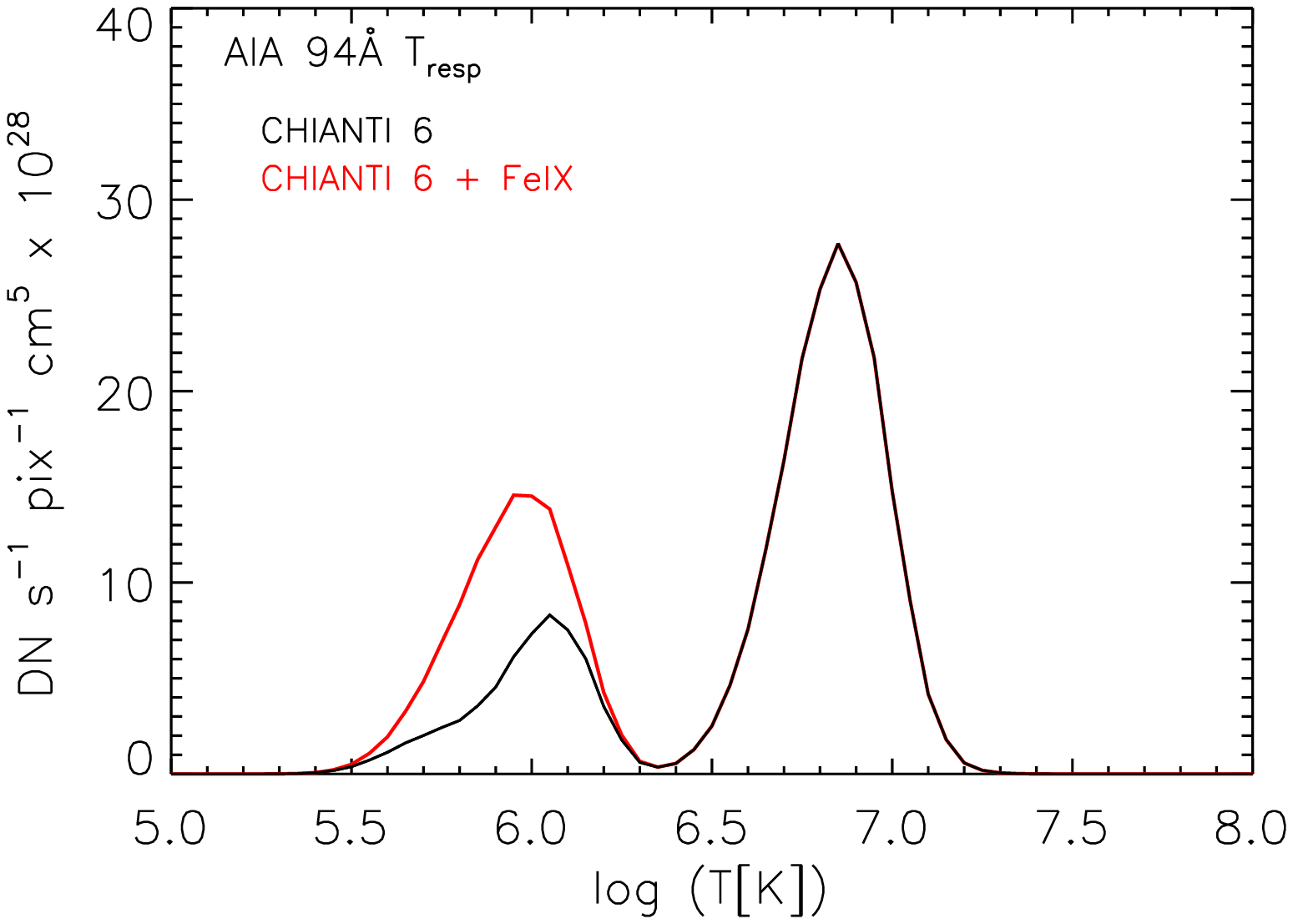}
\vspace{-0.3cm}}
\caption{Temperature response of AIA in the 94\AA\ passband: the black
  line represents the default response, obtained using the CHIANTI
  database (v.6.0.1), ionization equilibrium by \cite{Bryans09} and
  solar coronal abundances \citep{Feldman92}; the red curve represents the
  response including the contribution of \feix\ lines (Foster \& Testa, 2011) 
  not included in CHIANTI but observed in laboratory experiments
  \citep{Lepson02}. 
	\label{fig:resp_mod_emp}}
\end{figure}

If we include the contribution functions of these \feix\
lines, we can estimate the impact on the temperature response of the
AIA 94\AA\ passband. The ``corrected'' AIA temperature response is
shown in Figure~\ref{fig:resp_mod_emp}, compared to the default
response computed using CHIANTI 6.
Addition of the \feix\ contribution increases the response of the AIA 
94\AA\ channel by roughly a factor 2 in the cool temperature range
(see also Foster \& Testa 2011, where APED is used; \citealt{Smith01}). 

In the 131\AA\ wavelength region, none of the lines identified in 
\cite{Lepson02} appear to have significant impact for AIA observations.
However, some might be relevant to the analysis of EVE spectra, such as 
for instance two \feix\ $3p^54f \rightarrow 3p^53d$ transitions 
with measured wavelengths of  $\lambda$ 134.08\AA\ and 136.70\AA\ 
(see discussion in Foster \& Testa 2011, where a synthetic spectrum 
including these transitions is compared with a EVE quiet Sun spectrum).
Besides these lines around $\sim 94$\AA\ and $\sim 135$\AA, 
the new calculations presented in Foster \& Testa 2011 indicate that 
the added \feix\ transitions provide only limited additional flux in the 
10-170\AA\ band, compared to current versions of atomic databases. 
This modest additional emission is expected around 82\AA, in the 
110-115\AA\ range, and around $\sim 164$\AA\ where the model 
predicts a strong 4p-3d transition.

The study by \cite{Liang10} is also potentially helpful for identifying
lines missing in the atomic databases. They compare their \fevii-\fexvi\ 
atomic data with \cha/LETGS observations of Procyon. 
As they note, CHIANTI 6 only includes data from n=3 levels for 
\feix, \fexii, \fexiii, \fexiv, and the missing lines from these ions might
contribute non-negligible flux in the 10-170\AA\ wavelength range.
Although we use a set of spectral observations partially 
overlapping with those used by \cite{Liang10}, their aim and 
methods were different with respect to ours. 
They carry out a detailed comparison of their theoretical models with 
existing alternative atomic data focusing exclusively on \fevii-\fexvi\ lines, and 
they did not analyze the spectra longward of 106\AA.
Also, they used a simplified 3 temperature model from \citet{Raassen02} which
is expected to be a less accurate representation of the temperature distribution
of the coronal plasma compared to the emission measure distribution we use here 
and therefore lead to a less accurate reproduction of the line fluxes in the whole
wavelength range (see also above comparison of our finding for the EM(T) with 
the results of \citealt{Raassen02}).  Finally, the total exposure time of the 
observations we analyze here is almost twice as long as the
one used by \cite{Liang10}, improving the signal-to-noise ratio
of the spectrum.
For the 94\AA\ wavelength range, they noted problems in their model for 
\fex, which is predicting the 94.02\AA\ flux to be much lower than the 
CHIANTI 6 expected flux (and much lower than the observed flux). 
They did not include the \feix\ lines around 
$\sim 94$\AA\ observed in the EBIT experiments \citep{Lepson02}.

We note that recent analyses of AIA data have highlighted issues with 
the observed fluxes in the 94\AA\ and 131\AA\ passbands, suggesting
that (cool) lines missing in the atomic databases are the likely cause 
\citep{Schmelz11,Aschwanden11}. \cite{Aschwanden11} also attempted 
an estimate of the correction factor for the response of the 94\AA\ AIA 
channel to the cool (1MK) plasma, by analyzing a sample of loops for which 
emission in the other AIA bands is compatible with a near isothermal 
emission measure distribution. They derive a factor of $6.7 \pm 1.7$, which
is significantly larger than the contribution of \feix\ estimated by
\cite{Foster11} and in this work, suggesting that other ions 
also provide non-negligible contributions 
\footnote{Boerner et al. (in preparation) by looking at the morphology of
quiet Sun in deep exposures in the AIA passbands suggest possible 
additional contribution of emission from $\log T[K] \sim 6.0-6.3$.}.

In the 131\AA\ range the EBIT experiments suggest the presence of several 
\fevii\ lines, in the 127-134\AA\ range, which are not included in CHIANTI 
6.0.1 (or APED) and might provide significant contribution. Also a \feix\ 
line (134.08\AA) and a \fex\ line (134.09\AA) are observed in the EBIT spectra,
though their contribution might be rather limited given that the AIA response 
at those wavelenghts is already roughly two orders 
of magnitudes lower than at its peak.
Additional contribution can also come from ions other than the \fevii-\fex\ 
studied in the EBIT laboratory experiments.
In particular L-shell transitions of Mg and Ne are expected to fall around 94\AA\ 
and 132\AA. 
At the time of writing, the new version of the CHIANTI database (v7) 
is still under testing, though based on the changes in the ions relevant to 
the AIA bandpasses under consideration no significant differences are 
expected.

\section{Conclusions}
\label{s:conclusions}

High-resolution X-ray spectra of the low-activity
solar-like corona of Procyon obtained with the \cha\ Low Energy
Transmission Grating Spectrometer (LETGS) have been used for 
testing the X-ray/EUV data in the CHIANTI database. A systematic
benchmark ion by ion will be presented in two future papers (Drake
et al. 2011, and Landi et al., 2011, in preparation).
Model and observed spectra are in reasonably 
good agreement in the soft X-ray range ($\lambda \lesssim 50$\AA) and
at the longer LETGS wavelengths $\lambda \gtrsim 130$\AA.
However, the model flux lies significantly below the observed flux in the 
50-130\AA\ wavelength range.  In particular, in the 94\AA\ and 131\AA\  
AIA bands the observed flux exceeds the model flux by factors of $\sim 3$ 
and $\sim 1.9$, respectively.  By including two relatively strong \feix\ 
lines at $\lambda$ 93.59\AA\ and 94.07\AA\ 
observed in the laboratory by \citet{Lepson02}, the discrepancy in the 
94\AA\ band is reduced to $\sim 80$\%.
The AIA temperature response corrected in this way is
increased by roughly a factor of 2 at $10^6$~K.   
In the 131\AA\ band, \fevii\ transitions, not included in CHIANTI but observed 
in laboratory experiments, as well as L-shell Ne transitions, might explain
part of the missing flux.

\begin{acknowledgements}
We thank Adam Foster for useful discussion, and the referee for comments
that have helped improve the paper. 
PT was supported by contract SP02H1701R from Lockheed-Martin to the
Smithsonian Astrophysical Observatory. JJD was supported by NASA 
contract NAS8-39073 to the {\em Chandra X-ray Center} and thanks the 
Director, H. Tananbaum, for continuing support.
The work of EL is supported by the NNX11AC20G and NNX10AM17G 
grants to the University of Michigan, and grant SV1-81002 to the 
Smithsonian Astrophysical Observatory.
\end{acknowledgements}

\end{document}